\documentclass[12pt]{article}
\usepackage[colorlinks,linkcolor=Blue,citecolor=Blue,bookmarks,bookmarksnumbered]{hyperref}
\usepackage[scaled=0.85]{helvet}
\usepackage{XoohmE,accents}
\usepackage{graphicx,color}
\usepackage{booktabs}
\usepackage{multirow}
\usepackage{placeins}

\def\Dt#1{\accentset{\hbox{\LARGE.}}{#1}}	% dot-over for sp/sb
\def\DDt#1{\accentset{\hbox{\LARGE.\kern-2pt.}}{#1}}	% dot-over for sp/sb
\def\dt#1{\accentset{\hbox{\large.}}{#1}}	% dot-over for sp/sb
\def\ddt#1{\accentset{\hbox{\large\kern.5pt.\kern-1pt.}}{#1}}	% dot-over for sp/sb
\def\cb#1#2#3{\setlength\fboxsep{1pt}\colorbox{#1}{\color{#1}\fbox{\color{#2}#3}}}
\def\cB#1{\hbox to0pt{\setlength\fboxsep{0pt}\setlength\fboxrule{.3pt}\hss\color{grey3}%
           \fbox{\cb{white}{black}{#1}}\hss}}
\def\CB#1{\hbox to0pt{\setlength\fboxsep{0pt}\setlength\fboxrule{.3pt}\hss\color{black}%
           \fbox{\cb{black}{white}{#1}}\hss}}

\def\CLS{\text{\slshape CLS}}
\def\RSS{\text{\slshape RSS}}
\def\RPS{\text{\slshape RPS}}
\def\eX{\rlap{\raisebox{.35ex}{\kern.45ex\scriptsize\it=}}{\boldsymbol X}}
\def\eY{\rlap{\raisebox{.35ex}{\kern.12ex\scriptsize\it=}}{\boldsymbol Y}}
\def\EX{\rlap{\raisebox{.45ex}{\kern.425ex\scriptsize\it=}}{X}}
\def\EY{\rlap{\raisebox{.45ex}{\kern.1ex\scriptsize\it=}}{Y}}

\def\ad{{\dt\a}}
\def\bd{{\dt\b}}

\let\bs=\boldsymbol
\def\rD{{\rm D}}
\def\bD{\boldsymbol{D}\mkern-1.5mu}
\def\bDb{\hbox{\kern2pt\vrule height10pt depth-9.2pt width6pt\kern-9pt{$\boldsymbol D$}}\mkern-2mu}
\def\bQ{\boldsymbol{Q}}
\def\bQb{\hbox{\kern2pt\vrule height10pt depth-9.2pt width6pt\kern-9pt{$\boldsymbol Q$}}}

\def\bF{{\boldsymbol\Phi}}
\def\bJ{{\boldsymbol\Psi}}

\def\BSb{\hbox{\kern2.5pt\vrule height10pt depth-9.2pt width7pt\kern-10.25pt{$\boldsymbol{\mit\Sigma}$}}}

\def\rQb{\hbox{\kern1pt\vrule height10pt depth-9.2pt width6pt\kern-8pt{\bf Q}}}

\def\vC#1{\vcenter{\hbox{\hss#1\hss}}}
\def\Bm#1{\left[\begin{smallmatrix}#1\end{smallmatrix}\right]}
\def\rBx#1#2{\hbox to#1{#2\hss}}

\def\bX{{\bs{X}}}

\def\bY{{\bs{Y}\!}}

\def\vdt{\partial_\tau}

\definecolor{Hey}{rgb}{.9,.05,.4}
\definecolor{plum}{rgb}{.4,0,.6}

\definecolor{Green}  {rgb}{0.10,0.70,0.10} %  1
\definecolor{Orange} {rgb}{1.00,0.50,0.15} %  2
\definecolor{Red}    {rgb}{0.90,0.00,0.12} %  3
\definecolor{Purple} {rgb}{0.50,0.25,0.55} %  4
\definecolor{Turque} {rgb}{0.00,0.65,0.85} %  5
\definecolor{Blue}   {rgb}{0.00,0.00,1.00} %  6
\definecolor{Magenta}{rgb}{1.00,0.00,1.00} %  7
\definecolor{Gold}   {rgb}{1.00,0.75,0.25} %  8
\definecolor{Seaweed}{rgb}{0.01,0.24,0.09} %  9
\definecolor{Brown}  {rgb}{0.43,0.26,0.32} % 10
\definecolor{grey1}  {rgb}{0.20,0.20,0.20} % 11
\definecolor{grey2}  {rgb}{0.40,0.40,0.40} % 12
\definecolor{grey3}  {rgb}{0.60,0.60,0.60} % 13
\definecolor{grey4}  {rgb}{0.80,0.80,0.80} % 14
\definecolor{grey5}  {rgb}{0.90,0.90,0.90} % 15
\def\C#1#2{{\ifcase#1\or%Greg's color scheme
             \color{Green}\or \color{Orange}\or \color{Red}\or
              \color{Purple}\or \color{Turque}\or \color{Blue}\or
               \color{Magenta}\or \color{Gold}\or \color{Seaweed}\or
                \color{Brown}\or\color{grey1}\or\color{grey2}\or
                 \color{grey3}\else\color{grey4}\fi#2}}

\definecolor{Slate} {rgb}{0.00,0.45,0.55}
\def\Gr#1{{\color{Slate}#1}}

\newdimen\parshift\parshift=\parindent
\catcode`@=11
 \long\def\@footnotetext#1{\insert\footins{\reset@font\footnotesize
           \interlinepenalty\interfootnotelinepenalty\splittopskip%
            \footnotesep\splitmaxdepth\dp\strutbox\floatingpenalty\@MM%
             \hsize\columnwidth\addtolength{\hsize}{-2\parindent}
              \@parboxrestore\protected@edef\@currentlabel%
              {\csname p@footnote\endcsname\@thefnmark}%
                \color@begingroup%
                 \@makefntext{\rule\z@\footnotesep\ignorespaces#1%
                  \@finalstrut\strutbox}%
                \color@endgroup}}
 \long\def\@makefntext#1{\hglue\parshift%
           \vbox{\noindent\baselineskip=11pt plus.5pt minus.5pt\hb@xt@0em{\hss\@makefnmark\kern1pt}#1}}
\catcode`@=12
 \font\rOpe=cmsy10                        % Ersatz for the non-standard rope font
 \def\ktl{{\hbox{\rOpe\char'170}}}        % top end for left-handed rope
 \def\kbl{{\hbox{\rOpe\char'170}}}        % bottom end for left-handed rope
 \def\kcr{{\reflectbox{\rOpe\char'170}}}        % right-handed rope
 \def\ktr{{\reflectbox{\rOpe\char'170}}}        % top end for right-handed rope
 \def\kbr{{\reflectbox{\rOpe\char'170}}}        % bottom end for right-handed rope
 \def\Border{\vbox{\hsize0pt% braided border
        \setlength{\unitlength}{1mm}
        \newcount\xco
        \newcount\yco
        \xco=-21
        \yco=12
        \begin{picture}(0,0)(-7.5,0)
        \put(\xco,\yco){$\ktl$}
        \advance\yco by-1
        {\loop
        \put(\xco,\yco){$\kcr$}
        \advance\yco by-2
        \ifnum\yco>-240
        \repeat
        \put(\xco,\yco){$\kbl$}}
        \xco=170
        \yco=12
        \put(\xco,\yco){$\ktr$}
        \advance\yco by-1
        {\loop
        \put(\xco,\yco){$\kcr$}
        \advance\yco by-2
        \ifnum\yco>-240
        \repeat
        \put(\xco,\yco){$\kbr$}}
        \put(-19.5,13){\scalebox{.6065}{%
         University of Maryland Center for String and Particle  Theory \&\ Physics Department%
        |University of Maryland Center for String and Particle  Theory \&\ Physics Department}}
        \put(-19.5,-241.5){\scalebox{.5835}{%
         Howard University Department of Physics and Astronomy%
        |Howard University Department of Physics and Astronomy%
        |Howard University Department of Physics and Astronomy}}
        \end{picture}
        \par\vskip-8mm}}
\definecolor{UMred}{rgb}{.9,.05,.2}
\definecolor{HUblue}{rgb}{.0,.3,.7}
 \def\UMbanner{\vbox{\hsize0pt% The "UM" banner
        \setlength{\unitlength}{.4mm}
        \thicklines\color{UMred}
        \begin{picture}(0,0)(-30,-10)
        \put(165,16){\line(1,0){4}}
        \put(170,16){\line(1,0){4}}
        \put(180,16){\line(1,0){4}}
        \put(175,0){\line(1,0){4}}
        \put(180,0){\line(1,0){4}}
        \put(185,0){\line(1,0){4}}
        \put(169,0){\line(0,1){16}}
        \put(170,0){\line(0,1){16}}
        \put(179,0){\line(0,1){16}}
        \put(180,0){\line(0,1){16}}
        \put(184,0){\line(0,1){16}}
        \put(185,0){\line(0,1){16}}
        \put(169,16){\oval(8,32)[bl]}
        \put(170,16){\oval(8,32)[br]}
        \put(179,0){\oval(8,32)[tl]}
        \put(185,0){\oval(8,32)[tr]}
        \color{HUblue}
        \put(167.75,-2){\line(1,0){4}}
        \put(172.75,-2){\line(1,0){4}}
        \put(177.75,-2){\line(1,0){4}}
        \put(182.75,-2){\line(1,0){4}}
        \put(167.75,-2){\line(0,-1){16}}
        \put(171.75,-2){\line(0,-1){16}}
        \put(172.75,-2){\line(0,-1){16}}
        \put(176.75,-2){\line(0,-1){16}}
        \put(181.75,-2){\line(0,-1){16}}
        \put(182.75,-2){\line(0,-1){16}}
        \put(181.75,-2){\oval(8,32)[bl]}
        \put(182.75,-2){\oval(8,32)[br]}
        \put(167.75,-18){\line(1,0){4}}
        \put(172.75,-18){\line(1,0){4}}
        \end{picture}
        \par\vskip-6.5mm
        \thicklines}}

 \SfTitles %section titles more modest
 \allowdisplaybreaks
 \seceq

\begin{document}
%:Jim's border on the title page 
\thispagestyle{empty}
\vbox{\Border\UMbanner}
 \noindent{\small
 \today\hfill{UMDEPP 12-003 % un-comment-out and specify when done
 }}
  \vspace*{5mm}
 \begin{center}
{\LARGE\sf\bfseries The Real Anatomy of Complex Linear Superfields}\\[5mm]
  %\vfill
{\large\sf\bfseries S.J.\,Gates, Jr.$^*$,~
                     J. Hallett${}^{**}$,~
                    T.\,H\"{u}bsch$^{\dag\ddag}$~ and~
                    K.\,Stiffler$^*$
                    }\\*[4mm]
\emph{
      \centering
      $^*$Center for String and Particle Theory, Dept.\ of Physics, \\ University of Maryland, College Park, MD 20472,% {\tt  gatess@wam.umd.edu}
      \\\vspace{1.0mm}
      ${}^{**}$Department of Mathematics, Williams College, Williamstown, MA 01267 USA   
   \\\vspace{1.0mm}
     $^\ddag$Dept.\ of Physics, University of Central Florida, Orlando, FL 
     \\\vspace{1.0mm}
      $^\dag$Dept.\ of Physics \&\ Astronomy, Howard University, Washington, DC 20059%, {\tt  thubsch@howard.edu}
}
 \\\vspace*{2mm}
{\sf\bfseries ABSTRACT}\\[2mm]
\parbox{150mm}{\parindent=2pc\noindent\baselineskip=13pt plus1pt
Recent work on classification of off-shell representations of $N$-extended worldline supersymmetry without central charges has uncovered an unexpectedly vast number|trillions of even just (chromo)topology types|of so called {\em\/adinkraic\/} supermultiplets. Herein, we show by explicit analysis that a long-known but rarely used representation, the complex linear supermultiplet, is not adinkraic, cannot be decomposed locally, but may be reduced by means of a Wess-Zumino type gauge. This then indicates that the already unexpectedly vast number of adinkraic off-shell supersymmetry representations is but the proverbial tip of the iceberg.
 }
 
\end{center}
  %\vfill

%

%\newpage
\section{Introduction, Results and Summary}
 \label{IRS}
\subsection{History and Organization}
Refs.\cite{rA,r6-1,r6--1,r6-3,r6-3.2,r6-1.2} developed a detailed classification of a huge class (${\sim}\,10^{12}$ for no more than 32 supersymmetries) of {\em\/adinkraic\/} off-shell worldline supermultiplets, wherein each supercharge maps each component field to precisely one other component field or its $\t$-derivative. These supermultiplets are faithfully depicted by graphs called {\em\/Adinkras\/}; see also Refs.\cite{rPT,rT01,rT01a,rCRT,rKRT,rKT07,rGKT10}. That a basis of supercharges and a compatible basis of component fields that exhibits such an acute syzygy can even exist is by no means guaranteed. Herein, we demonstrate that an oft-ignored ``red-headed stepchild'' in the family of famous and familiar supermultiplets, the {\em\/complex linear supermultiplet\/}\cite{r1001}\ft{This was also called the {\em\/non-minimal chiral\/} supermultiplet in Ref.\cite{rHSS}.}|\CLS, for short, in fact is not adinkraic. This, in fact, {\em\/had\/} to be the case, since the 12+12 component fields of \CLS\ cannot possibly span a hypercubical chromotopology, as required by Theorem~4.1 of Ref.\cite{r6-3}.

In the remainder of this introduction, we specify the notation and conventions.  Section~\ref{s:RSMRPM} analyzes the real scalar supermultiplet (\RSS) and real pseudoscalar supermultiplet (\RPS) in their real/Majorana structure in 3+1-dimensional spacetime and reduces them to their worldlines, \ie, 0-brane. Section~\ref{s:CLM} merges the \RSS\ and \RPS\ into the \CLS\ and analyzes the resulting 3+1-dimensional spacetime structure and 0-brane reduction, following a similar pattern to Section~\ref{s:RSMRPM}. Section~\ref{s:End} presents the supersymmetric action for the \RSS, \RPS, and \CLS\ and concludes with a discussion of the implications of this result. 
%Some technically involved details of our analysis are deferred to the appendices.

\subsection{Definitions}
 \label{s:DDefs}
Complex linear superfields are defined by the property that they are annihilated by a Lorentz-invariant ``square'' of the conjugate superderivatives. We focus on the most familiar case, of simple supersymmetry in $1+3$-dimensional spacetime that has a total of $N=4$ supercharges. We consider the worldline (0-brane) dimensional reduction of this particle, and so operate with the $(1|4)$-supersymmetry algebra
\begin{equation}
 \left.
 \begin{aligned}
 \big\{\, Q_I \,,\, Q_J \,\big\}&=2\d_{IJ}\,H,&
 \big[\, H \,,\, Q_I \,\big] &=0,\\
 Q_I^{~\dagger}&= Q_I,& H^\dagger&=H,\quad
\end{aligned}\right\}\quad
 I,J=1,\cdots,4,
  \label{e:RSuSy}
\end{equation}
were $H=i\hbar\vdt$.
 We also have the superderivatives:
\begin{equation}
 \left.
 \begin{aligned}
 \big\{\, \rD_I \,,\, \rD_J \,\big\}&=2\d_{IJ}\,H,&
 \big[\, H \,,\, \rD_I \,\big] &=0,\\
 \rD_I^{~\dagger}&=-\rD_I,&
 \big\{\, Q_I \,,\, \rD_J \,\big\}&=0,\quad
\end{aligned}\right\}\quad
 I,J=1,\cdots,4.
  \label{e:RDuSy}
\end{equation}
Instead of this real basis, we may also use the complex (and nilpotent) basis of Hermitian pairs:
\begin{equation}
 \left.
 \begin{aligned}
 \bQ_\a&:=\inv2\big(\,Q_\a+iQ_{2+\a}\,\big),&
 \bQb_\ad&:=\bQ_\a^{~\dagger}
                =\inv2\big(\,Q_\ad-iQ_{2+\ad}\,\big),\\[1mm]
 \bD_\a&:=\inv2\big(\,\rD_\a+i\rD_{2+\a}\,\big),&
 \bDb_\ad&:=\bD_\a^{~\dagger}
                =\inv2\big(\,\rD_\ad-i\rD_{2+\ad}\,\big),\quad
\end{aligned}\right\}\quad
 \a,\Dt\a=1,2.
 \label{e:rDrQ}
\end{equation}
With this basis, the algebra\eq{e:RSuSy} and\eq{e:RDuSy} becomes:
 \begin{equation}
 \big\{\, \bQ_\a \,,\, \bQb_\bd \,\big\}=\d_{\a\bd}\,H,\qquad
 \big\{\, \bD_\a \,,\, \bDb_\bd \,\big\}=\d_{\a\bd}\,H,
\end{equation}
with all other (anti)commutators vanishing. In particular, note that
\begin{equation}
 \bQ_\a^{~2}~=~0~=~\bQb_\bd^{~2},\quad\text{and}\quad
 \bD_\a^{~2}~=~0~=~\bDb_\bd^{~2}. \label{e:Nilpot}
\end{equation}
We will also need that
\begin{equation}
  Q_I=i\rD_I-2i\d_{IJ}\q^J\,H,\quad\text{and}\quad
  \rD_I=-iQ_I-2\d_{IJ}\q^J\,H,
 \label{e:Q=iD}
\end{equation}
where $\q^I$ provide the fermionic extension to (space)time into superspace. When applied on superfields (general functions over superspace), the $\rD_I$ act as left-derivatives while the $Q_I$ act as right-derivatives. Owing to this and with the sign conventions and definitions from Ref.\cite{rFGH}, we have that
\begin{equation}
  Q_I\,\f=i\rD_I\,\bs\F|,\qquad\text{and}\qquad
  Q_I\,\j=-i\rD_I\,\bs\J|,
 \label{e:Q=sD}
\end{equation}
where $\bs\F|=\f$ ($\bs\J|=\j$) is an arbitrary bosonic (fermionic) functional-differential expression, and $\bs\F$ ($\bs\J$) the appropriate superfield expression defining $\f$ ($\j$) by means of the superspace\,$\to$\,spacetime projection denoted by the ``$|$'' right-delimiter. Owing to the relations\eq{e:Q=sD}, a supermultiplet may be represented {\em\/interchangeably\/}:
\begin{subequations}
 \label{e:SM}
\begin{alignat}9
\text{as}&\quad
  \big\{\,\big(\bF\mid\bJ_I\mid{\cdots}\big):~&
          \rD_I\,\bF&=i\bJ_I,~& \rD_I\,\bJ_J&=\d_{IJ}\dt\bF+\dots,~& \etc\big\},
 \label{e:DSM}\\
\text{as}&\quad
  \big\{\,\big(\,\f\mid\,\j_I\mid\cdots\big):~&
          Q_I(\f)&=-\j_I,~& Q_I(\j_J)&=-i\d_{IJ}\dt\f+\dots,~& \etc\big\},
 \label{e:QSM}\\
\text{or as}&\quad
  \big\{\,\big(\,\f\mid\,\j_I\mid\cdots\big):~&
          \d_e(\f)&=i\e^I\j_I,~& \d_e(\j_J)&=-\e^I\d_{IJ}\dt\f+\dots,~& \etc\big\}.
 \label{e:dSM}
\end{alignat}
\end{subequations}
Relying on the\eqs{e:DSM}{e:dSM} correspondence, we may also specify supermultiplets in the latter representation, \eq{e:DSM}, as a manifestly supersymmetric (recall: $\{Q_I,\rD_J\}=0$) closed system of superdifferential relations between component superfields, the lowest component of each of which is the corresponding (ordinary spacetime) component field.

We will therefore write our transformation laws in component notation in terms of the superderivative ${\rm D}_I$. On way to our various 0-brane reductions, we will always start with $4D$,  ${\mathcal N} =1$ transformation laws written in terms of ${ \rm D}_a $ with $a=1,2,3,4$ a Majorana fermionic index. Throughout the paper, we denote the number of supersymmetries on the 0-brane as $N = 4~{\mathcal N}$, where ${\mathcal N}$ is always the number of supersymmetries in $4D$. We will use the real representation of the $\gamma$ matrices as in Refs.\cite{rUMD09-1,rUMD09-2}
\begin{align}
  (\gamma^0)_{a}^{~b} = &\left(
\begin{array}{cccc}
 0 & 1 & 0 & 0 \\
 -1 & 0 & 0 & 0 \\
 0 & 0 & 0 & -1 \\
 0 & 0 & 1 & 0
\end{array}
\right) ~~~,~~~(\gamma^1)_a^{~b} = \left(
\begin{array}{cccc}
 0 & 1 & 0 & 0 \\
 1 & 0 & 0 & 0 \\
 0 & 0 & 0 & 1 \\
 0 & 0 & 1 & 0
\end{array}
\right)\cr
  (\gamma^2)_{a}^{~b} = & \left(
\begin{array}{cccc}
 0 & 0 & 0 & -1 \\
 0 & 0 & 1 & 0 \\
 0 & 1 & 0 & 0 \\
 -1 & 0 & 0 & 0
\end{array}
\right)~~~,~~~(\gamma^3)_a^{~b} = \left(
\begin{array}{cccc}
 1 & 0 & 0 & 0 \\
 0 & -1 & 0 & 0 \\
 0 & 0 & 1 & 0 \\
 0 & 0 & 0 & -1
\end{array}
\right)\\
  (\gamma^5)_{a}^{~b} \equiv & i (\gamma^0 \gamma^1 \gamma^2 \gamma^3)_{a}^{~b} = \left(
\begin{array}{cccc}
 0 & 0 & 0 & i \\
 0 & 0 & -i & 0 \\
 0 & i & 0 & 0 \\
 -i & 0 & 0 & 0
\end{array}
\right)
\end{align} 
where the fermionic indices are raised and lowered by the spinor metric $C_{ab}$ and inverse
 spinor metric $C^{ab}$
\begin{align}
C_{ab} = & \left(
\begin{array}{cccc}
 0 & -1 & 0 & 0 \\
 1 & 0 & 0 & 0 \\
 0 & 0 & 0 & 1 \\
 0 & 0 & -1 & 0
\end{array}
\right)~~~,~~~C^{ab} = \left(
\begin{array}{cccc}
 0 & -1 & 0 & 0 \\
 1 & 0 & 0 & 0 \\
 0 & 0 & 0 & 1 \\
 0 & 0 & -1 & 0
\end{array}
\right)
\end{align}
according to the convention
\begin{align}
    \bs \zeta_a = \bs \zeta^b C_{ba}~~~,~~~\bs \zeta^a = C^{ab} \bs \zeta_b ~~~,
\end{align}
which satisfy
\begin{align}
  C_{ac} C^{bc} = \delta_a^{~b} ~~~.
\end{align}

\section{The Real Scalar and Pseudoscalar Supermultiplets}\label{s:RSMRPM}
In this section we will introduce both the real scalar and real pseudoscalar supermultiplets in our $4D$, ${\mathcal N} = 1$ component notation. We will then dimensionally reduce these multiplets to the 0-brane by considering all fields to have only time dependence. Finally, we will organize these dimensionally reduced transformation laws into Adinkras for both supermultiplets.

\subsection{\texorpdfstring{$4D$}{4D}, \texorpdfstring{${\mathcal N} = 1$}{N=1} Transformation Laws}
In component form, the real scalar supermultiplet can be written as the following set of supersymmetric transformation laws
\begin{subequations}
\label{e:DRSM}
\begin{align}
 { \rm D}_a \bs K= & \bs \z_a \\
 %%%%%%%%%%%%%%%%%%%%%%%%%%
  { \rm D}_a \bs \z_b = & i (\g^\mu)_{ab} \partial_\mu \bs K + ( \g^5 \g^\mu)_{ab} {\bs 
  U}_\mu +i C_{ab} \bs M + (\g^5)_{ab} \bs N \\
  %%%%%%%%%%%%%%%%%%%%%%%%%%
 { \rm D}_a \bs M = & \frc{1}2\bs \L_a -\frc{1}2 (\g^\nu)_{a}^{\,\,\,d} \partial_\nu \bs \z_d \\
  %%%%%%%%%%%%%%%%%%%%%%%%%%
 { \rm D}_a \bs N = &-i \frc{1}2 (\g^5)_a^{\,\,\,\,d} \bs \L_d +i \frc{1}2 (\g^5 \g^\nu)_{a}^{\,\,\,d} 
 \partial_\nu \bs \z_d \\
  %%%%%%%%%%%%%%%%%%%%%%%%%%
 { \rm D}_a \bs U_\mu = & i \frc{1}2(\g^5 \g_\mu)_a^{\,\,\,\,d} \bs \L_d -i \frc{1}2 (\g^5 \g^\nu 
 \g_\mu)_{a}^{\,\,\,d} 
 \partial_\nu \bs \z_d \\
  %%%%%%%%%%%%%%%%%%%%%%%%%%
 { \rm D}_a \bs \L_b = & i (\g^\mu)_{ab} \partial_\mu \bs M + (\g^5 \g^\mu)_{ab} \partial_\mu 
 \bs N+ ( \g^5 \g^\mu  \g^\nu)_{ab} \partial_\mu \bs U_\nu + i C_{ab} \bs {\rm d} \\
  %%%%%%%%%%%%%%%%%%%%%%%%%%
 { \rm D}_a \bs {\rm d} = & -(\g^\nu)_a^{\,\,\,d} \partial_\nu \bs \L_d. 
\end{align}
\end{subequations}
where ${\rm D}_a$ is the superderivative alluded to in Section~\ref{IRS}.  The superderivatives satisfy the closure relation
\begin{align}\label{e:DaDbclosure}
   \{ {\rm D}_a, {\rm D}_b \} = 2 i (\gamma^\mu)_{ab} \partial_\mu
\end{align}
for all fields in the real scalar supermultiplet\eq{e:DRSM}.

The real pseudoscalar supermultiplet can be defined from the real scalar supermultiplet via
\begin{align}\label{substitution1}
\bs K\rightarrow  \bs L&,~~~ \bs \zeta_a\rightarrow 
i (\gamma^5)_{a}^{\,\,\,d}\bs \rho_d,~~~ \bs M\rightarrow\Tw{\bs N},~~~  \bs N\rightarrow 
-\Tw{\bs M}\cr
&\bs U_\mu\rightarrow \bs V_\mu,~~~ \bs\Lambda_a\rightarrow -i(\gamma^5)_{a}^{\,\,\,d}\Tw{\bs\Lambda}_d,~~~ \bs {\rm d}\rightarrow\Tw{\bs {\rm d}} 
\end{align}
and this yields the pseudoscalar supermultiplet (\RPS), which satisfies
\begin{subequations}\label{e:DRPM}
\begin{align}
{\rm D}_a\bs L= &i (\gamma^5)_{a}^{\,\,\,d}\bs \rho_d \\
%%%%%%%%%%%%%%%%%%%%%%%%%%%%%
{\rm D}_a\bs \rho_b= &-(\gamma^5\gamma^\mu)_{ab}\partial_\mu \bs L+i(\gamma^\mu
)_{ab}\bs V_\mu+(\gamma^5)_{ab}\Tw{\bs N}+iC_{ab}\Tw{\bs M} \\
%%%%%%%%%%%%%%%%%%%%%%%%%%%%%
{\rm D}_a\Tw{\bs N}=&-i \frc{1}2 (\gamma^5)_{a}^{\,\,\,d}\Tw{\bs \Lambda}_d+i\frc{1}2 
(\gamma^5\gamma^\mu)_{a}^{\,\,\,d}\partial_\mu\bs \rho_d \\
%%%%%%%%%%%%%%%%%%%%%%%%%%%%%
{\rm D}_a\Tw{\bs M}= &\frc{1}2 \Tw{\bs \Lambda}_a-\frc{1}2 (\gamma^\mu)_{a}^{\,\,\,d}\partial_\mu\bs \rho_d \\
%%%%%%%%%%%%%%%%%%%%%%%%%%%%%
{\rm D}_a \bs V_\mu= &- \frc{1}2 (\gamma_\mu)_{a}^{\,\,\,d}\Tw{\bs \Lambda}_d+\frc{1}2 (\gamma^\nu
\gamma_\mu)_{a}^{\,\,\,d}\partial_\nu\bs \rho_d \\
%%%%%%%%%%%%%%%%%%%%%%%%%%%%%%%%%%%%%%%%%%%%%%%%%%%%%%%%%%%%%%
{\rm D}_a\Tw{\bs \Lambda}_b=&(\gamma^5\gamma^\mu)_{ab}\partial_\mu\Tw{\bs N}+i (\gamma^\mu)_{ab}
\partial_\mu\Tw{\bs M}+i(\gamma^\mu\gamma^\nu)_{ab}\partial_\mu \bs V_\nu-(\gamma^5)_{ab}\Tw{\bs {\rm d}} \\
%%%%%%%%%%%%%%%%%%%%%%%%%%%%%
{\rm D}_a\Tw{\bs {\rm d}}= & -i (\gamma^5\gamma^\mu)_{a}^{\,\,\,d}\partial_\mu\Tw{\bs \Lambda}_d ~~.
\end{align}
\end{subequations}
where the superderivatives again satisfy the closure relation~\eq{e:DaDbclosure} for all fields in the real pseudoscalar supermultiplet\eq{e:DRPM}.

\subsection{Dimensional Reduction and Adinkras}
Performing the one-dimensional reduction of the real scalar supermultiplet\eq{e:DRSM} and defining $\bs X_{IJ} = - \bs X _{JI} \equiv -i\frc{1}2 {\rm D}_{[I} {\rm D}_{J]} \bs K =-i\frc{1}2({\rm D}_{I} {\rm D}_{J} - {\rm D}_{J} {\rm D}_{I} ) \bs K $ we find:
\begin{align}\label{e:RSSX}
    \bs X_{12} \equiv & \, \bs U_2 - \bs M,~~~\bs X_{13} \equiv \bs U_3 - \bs N,~~~\bs  X_{23} \equiv \bs U_1 - \bs U_0, \cr
     \bs X_{14} \equiv & \,  \bs U_0 + \bs U_1,~~~\bs X_{24} \equiv -\bs U_3 - \bs N,~~~\bs X_{34} \equiv \bs U_2 + \bs M
\end{align} we can write the transformation laws as in the Table~\ref{t:ScX}.
\begin{table}[htp]
\caption{The superdifferential relations between the one-dimensionally reduced components of the real scalar superfield.}
\small\vspace*{-4mm}
$$\begin{array}{@{} c@{:~}c|c@{~}c@{~}c@{~}c|c@{~}c@{~}c@{~}c@{~}c@{~}c|c@{~}c@{~}c@{~}c|c @{}}
 \omit& \bs K &\bs \zeta_1 &\bs \zeta_2 &\bs \zeta_3 &\bs \zeta_4
      &\bX_{12} &\bX_{13} &\bX_{14} &\bX_{23} &\bX_{24} &\bX_{34}
      &-\bs \Lambda_2 &\bs \Lambda_1 &\bs \Lambda_4 &-\bs \Lambda_3 &\bs {\rm d}\strut\\ 
    \toprule
\C3{\rD_1} &\bs \zeta_1 & i \Dt{\bs K} & i \bX_{12} & i \bX_{13} & i \bX_{14}
           &\Dt{\bs \zeta}_2 &\Dt{\bs \zeta}_3 &\Dt{\bs \zeta}_4 &-\bs \Lambda_3 &-\bs \Lambda_4 &\bs \Lambda_1 
           & i \bs {\rm d} & i \Dt{\bX}_{34} &-i\Dt{\bX}_{24} &i\Dt{\bX}_{23} &-\Dt{\bs \Lambda}_2 \\
\C1{\rD_2} &\bs \zeta_2 &-i\bX_{12} &i\Dt{\bs K} &i\bX_{23} &i\bX_{24}
           &-\Dt{\bs \zeta}_1 &\bs \Lambda_3 &\bs \Lambda_4 &\Dt{\bs \zeta}_3 &\Dt{\bs \zeta}_4 &\bs \Lambda_2
           &-i\Dt{\bX}_{34} &i\bs {\rm d} &i\Dt{\bX}_{14} &-i\Dt{\bX}_{13} &\Dt{\bs \Lambda}_1\\
\C6{\rD_3} &\bs \zeta_3 &-i\bX_{13} &-i\bX_{23} &i\Dt{\bs K} &i\bX_{34}
           &-\bs \Lambda_3 &-\Dt{\bs \zeta}_1 &-\bs \Lambda_1 &-\Dt{\bs \zeta}_2 & -\bs \Lambda_2 &\Dt{\bs \zeta}_4
           &i\Dt{\bX}_{24} &-i\Dt{\bX}_{14} &i\bs {\rm d} &i\Dt{\bX}_{12} &\Dt{\bs \Lambda}_4\\ 
\C2{\rD_4} &\bs \zeta_4 &-i\bX_{14} &-i\bX_{24} &-i\bX_{34} &i\Dt{\bs K}
           &-\bs \Lambda_4 &\bs \Lambda_1 &-\Dt{\bs \zeta}_1 &\bs \Lambda_2 &-\Dt{\bs \zeta}_2 &-\Dt{\bs \zeta}_3 
           &-i\Dt{\bX}_{23} &i\Dt{\bX}_{13} &-i\Dt{\bX}_{12} &i\bs {\rm d} &-\Dt{\bs \Lambda}_3\\ 
    \bottomrule
  \end{array}$$
\label{t:ScX}
\end{table}%
With the conventions of Refs.\cite{rUMD09-1,rUMD09-2}, we can depict the dimensionally reduced transformation laws for the real scalar supermultiplet  (\RSS) as the Adinkra in ~(\ref{e:X})

\begin{equation}\unitlength=.9mm
 \vC{\begin{picture}(160,55)
   \put(0,0){\includegraphics[width=146mm]{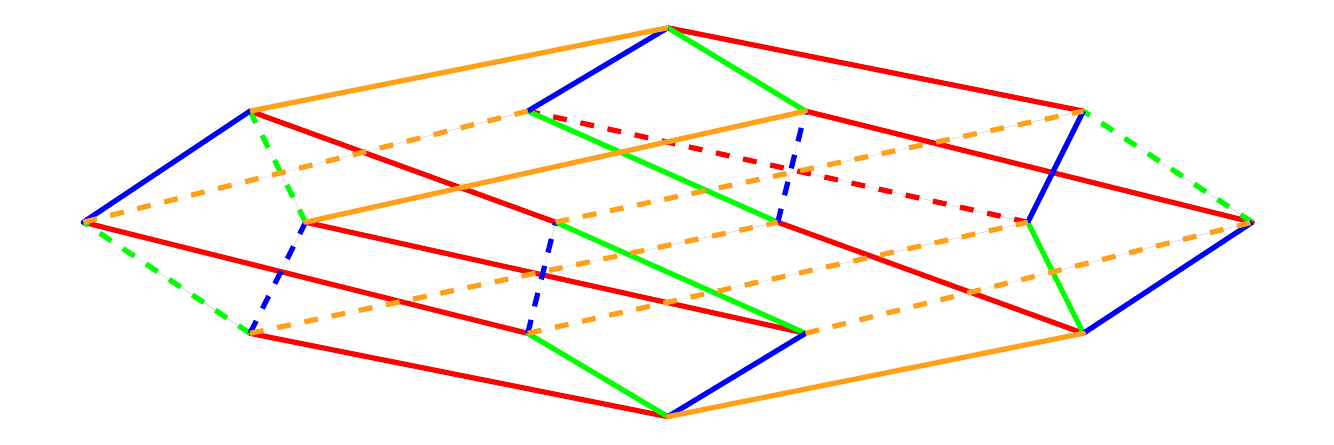}}
     \put(10,45){\mbox{\RSS:}}
     \put(80,49){\cB{$\bs {\rm d}$}}
     \put(30,38.5){\CB{$-\bs \Lambda_3$}}
     \put(64,38.5){\CB{$\bs \Lambda_4$}}
     \put(96,38.5){\CB{$\bs \Lambda_1$}}
     \put(131,38.5){\CB{$-\bs \Lambda_2$}}
     \put(10,25.5){\cB{$\bs X_{12}$}}
     \put(36,25.5){\cB{$\bs X_{13}$}}
     \put(68,25.5){\cB{$\bs X_{23}$}}
     \put(93,25.5){\cB{$\bs X_{14}$}}
     \put(124,25.5){\cB{$\bs X_{24}$}}
     \put(150,25.5){\cB{$\bX_{34}$}}
     \put(30,12.5){\CB{$\bs \zeta_1$}}
     \put(64,12.5){\CB{$\bs \zeta_2$}}
     \put(96,12.5){\CB{$\bs \zeta_3$}}
     \put(131,12.5){\CB{$\bs \zeta_4$}}
     \put(80,2.5){\cB{$\bs K$}}
 \end{picture}}
 \label{e:X}
\end{equation}
where \eg, $\C3{\rD_1}\,\bs K=\bs \zeta_1$  is depicted by the solid red edge, and $\C3{\rD_1}\,\bX_{24}=-\bs \Lambda_4$ by the dashed red edge where the dashing encodes the minus sign; following an edge downward results in the $\t$-derivative of the indicated component field. Owing to the factors of $i$ in \emph{all} fermion to boson transformations in Table~\ref{t:ScX}, the $\bs \zeta_2$\C3{\rule[.5ex]{10pt}{1pt}}$\bX_{12}$ solid red edge depicts for instance the relation $\C3{\rD_1}\,\bs \zeta_2=i \bX_{12}$.

With the substitutions ~(\ref{substitution1}), written more explicitly as in Eq.~(\ref{substitution1expanded})
\begin{align}\label{substitution1expanded}
\bs K\rightarrow  \bs L&,~~~ \bs M\rightarrow\Tw{\bs N},~~~  \bs N\rightarrow 
-\Tw{\bs M}\cr
&\bs U_\mu\rightarrow \bs V_\mu,~~~ \bs {\rm d}\rightarrow\Tw{\bs {\rm d}}  \cr
\bs \zeta_1\rightarrow -\bs \rho_4,~~~&\bs \zeta_2\rightarrow  \bs \rho_3,~~~\bs \zeta_3\rightarrow -\bs \rho_2,~~~\bs \zeta_4\rightarrow \bs \rho_1 \cr 
\bs\Lambda_1 \rightarrow \Tw{\bs\Lambda}_4,~~~\bs\Lambda_2 &\rightarrow -\Tw{\bs\Lambda}_3,~~~\bs\Lambda_3 \rightarrow \Tw{\bs\Lambda}_2,~~~\bs\Lambda_4 \rightarrow -\Tw{\bs\Lambda}_1,
\end{align}
and the definitions $\bs Y_{IJ} = - \bs Y_{JI} \equiv -i \frc{1}2{\rm D}_{[I} {\rm D}_{J]} \bs L$:
\begin{align}\label{e:RPSY}
  \bs Y_{12} \equiv & \bs V_2 - \Tw{ \bs N}~~~, \bs Y_{13} \equiv  \bs V_3 + \Tw{\bs M},~~~ \bs Y_{23} \equiv  \bs V_1 - \bs V_0 \cr
  \bs Y_{14} \equiv & \bs V_0 + \bs V_1,~~~ \bs Y_{24} \equiv \Tw{\bs M} - \bs V_3,~~~ \bs Y_{34} \equiv \bs V_2 + \Tw{\bs N}
\end{align}
Table~\ref{t:ScX} transforms into Table~\ref{t:ScY} for the real pseudoscalar supermultiplet\eq{e:DRPM} which can then be depicted succinctly as the Adinkra in~(\ref{e:Y}).
\begin{table}[htp]
\caption{The superdifferential relations between the one-dimensionally reduced components of the real pseudoscalar superfield.}
\small\vspace*{-4mm}
$$\begin{array}{@{} c@{:~}c|c@{~}c@{~}c@{~}c|c@{~}c@{~}c@{~}c@{~}c@{~}c|c@{~}c@{~}c@{~}c|c @{}}
 \omit& \bs L &-\bs \rho_4 &\bs \rho_3 &-\bs \rho_2 &\bs \rho_1
      &\bY_{12} &\bY_{13} &\bY_{14} &\bY_{23} &\bY_{24} &\bY_{34}
      &\Tw{\bs \Lambda}_3 &\Tw{\bs \Lambda}_4 &-\Tw{\bs \Lambda}_1 &-\Tw{\bs \Lambda}_2 &\Tw{\bs {\rm d}}\strut\\ 
    \toprule
\C3{\rD_1} &-\bs \rho_4 & i \Dt{\bs L} & i \bY_{12} & i \bY_{13} & i \bY_{14}
           &\Dt{\bs \rho}_3 &-\Dt{\bs \rho}_2 &\Dt{\bs \rho}_1 &-\Tw{\bs \Lambda}_2 &\Tw{\bs \Lambda}_1 &\Tw{\bs \Lambda}_4 
           & i \Tw{\bs {\rm d}} & i \Dt{\bY}_{34} &-i\Dt{\bY}_{24} &i\Dt{\bY}_{23} &\Dt{\Tw{\bs \Lambda}}_3 \\
\C1{\rD_2} &\bs \rho_3 &-i\bY_{12} &i\Dt{\bs L} &i\bY_{23} &i\bY_{24}
           &\Dt{\bs \rho}_4 &\Tw{\bs \Lambda}_2 &-\Tw{\bs \Lambda}_1 &-\Dt{\bs \rho}_2 &\Dt{\bs \rho}_1 &-\Tw{\bs \Lambda}_3
           &-i\Dt{\bY}_{34} &i\Tw{\bs {\rm d}} &i\Dt{\bY}_{14} &-i\Dt{\bY}_{13} &\Dt{\Tw{\bs \Lambda}}_4\\
\C6{\rD_3} &-\bs \rho_2 &-i\bY_{13} &-i\bY_{23} &i\Dt{\bs L} &i\bY_{34}
           &-\Tw{\bs \Lambda}_2 &\Dt{\bs \rho}_4 &-\Tw{\bs \Lambda}_4 &-\Dt{\bs \rho}_3 & \Tw{\bs \Lambda}_3 &\Dt{\bs \rho}_1
           &i\Dt{\bY}_{24} &-i\Dt{\bY}_{14} &i\Tw{\bs {\rm d}} &i\Dt{\bY}_{12} &-\Dt{\Tw{\bs \Lambda}}_1\\ 
\C2{\rD_4} &\bs \rho_1 &-i\bY_{14} &-i\bY_{24} &-i\bY_{34} &i\Dt{\bs L}
           &\Tw{\bs \Lambda}_1 &\Tw{\bs \Lambda}_4 &\Dt{\bs \rho}_4 &-\Tw{\bs \Lambda}_3 &-\Dt{\bs \rho}_3 &\Dt{\bs \rho}_2 
           &-i\Dt{\bY}_{23} &i\Dt{\bY}_{13} &-i\Dt{\bY}_{12} &i\Tw{\bs {\rm d}} &-\Dt{\Tw{\bs \Lambda}}_2\\ 
    \bottomrule
  \end{array}$$
\label{t:ScY}
\end{table}%
\begin{equation}\unitlength=.9mm
 \vC{\begin{picture}(160,55)
   \put(0,0){\includegraphics[width=146mm]{Spindle.pdf}}
     \put(10,45){\mbox{\RPS:}}
     \put(80,49){\cB{$\Tw{\bs {\rm d}}$}}
     \put(30,38.5){\CB{$-\Tw{\bs \Lambda}_2$}}
     \put(64,38.5){\CB{$-\Tw{\bs \Lambda}_1$}}
     \put(96,38.5){\CB{$\Tw{\bs \Lambda}_4$}}
     \put(131,38.5){\CB{$\Tw{\bs \Lambda}_3$}}
     \put(10,25.5){\cB{$\bs Y_{12}$}}
     \put(36,25.5){\cB{$\bs Y_{13}$}}
     \put(68,25.5){\cB{$\bs Y_{23}$}}
     \put(93,25.5){\cB{$\bs Y_{14}$}}
     \put(124,25.5){\cB{$\bs Y_{24}$}}
     \put(150,25.5){\cB{$\bs Y_{34}$}}
     \put(30,12.5){\CB{$-\bs \rho_4$}}
     \put(64,12.5){\CB{$\bs \rho_3$}}
     \put(96,12.5){\CB{$-\bs \rho_2$}}
     \put(131,12.5){\CB{$\bs \rho_1$}}
     \put(80,2.5){\cB{$\bs L$}}
 \end{picture}}
 \label{e:Y}
\end{equation}
\section{The Complex Linear Supermultiplet}\label{s:CLM}

\subsection{The Complex Linear Supermultiplet Fusion}
\label{s:RChiral}
In the introduction, the complex linear supermultiplet was called a ``red-headed stepchild.''  
The reason for this can be seen by reviewing its history. The proper appearance of the 
complex linear supermultiplet occurred in works on the subjects of superfield 
supergravity\cite{rWSJG79} and the use of superspace\cite{r1001}.  The first comment 
that it could be related by a duality transformation to the usual chiral multiplet was 
contained in Ref.\cite{rJGWS81} where it was noted that Zumino had previously given a 
first-order action appropriate for this purpose\cite{rBZ80}.
In 1984, it was proposed\cite{rBDJG85} that the existence of the chiral supermultiplet and the
complex linear supermultiplet offered the possibility of assigning left-handed and
right-handed fermions to very distinct supersymmetry representations with one handedness
being described by chiral supermultiplet and the other handedness to arise from the
complex linear supermultiplet.  Since that time, one of the authors (SJG) has returned
to investigate the complex linear supermultiplet many times in applications such as
phenomenological actions\cite{rJG96,rJG96-2}, six dimensional SUSY sigma-models \cite{rJGSPGM06,rJGSPGM06b},
4D $\cal N$ = 2 SUSY sigma-models\cite{rJGSK99,rJGTHSK99,rJGSK99b}, and in relation to a fundamental
representation theory\cite{r6-2,r6-4.2}.  In recent times, principally due to S.~Kuzenko and collaborators,
there have appeared many new and impressive results\cite{rSK06,rSK06b,rSKULRU07,rSK07,rMASKUL07,rMASKUL07b,rSKJN08,rSK09,rSKULRU09,rSKULRU10,rSK10,rDBSK11,rDBSK11b,rSK11,rDBSK11c,rSKST11} that lend broad and deep 
insights into the use of the complex linear multiplet in additional topics such as 
Goldstone superfields, sigma-models,  and their relation to harmonic and projective
superspace models.

So by this point, the ``red-headed stepchild'' has been shown to provide a wide 
array of applications.  There are even some hints\cite{rChiLin,rJLouis} that the complex linear supermultiplet may be of interest for effective string theories.
  
  We can construct the complex linear supermultiplet from fusion of a real scalar and pseudoscalar supermultiplet with the constraint
  \begin{align}\label{e:CLSprimary}
     (C^{ab} + (\gamma^5)^{ab}){\rm D}_a {\rm D}_b (\bs K + i \bs L) = 0.
  \end{align}
  As $C^{ab}$ is purely real and $(\gamma^5)^{ab}$ is purely imaginary, we can split this constraint into its real and imaginary parts 
\begin{subequations}
  \label{e:CLSprimaryRealImaginary}
  \begin{align}
       \label{e:CLSprimaryReal}
        {\rm D}_a {\rm D}_b (C^{ab}\bs K + i (\gamma^5)^{ab}\bs L) &= 0 ~~~\Rightarrow ~~~ ({\rm D}_{[3} {\rm D}_{4]} - {\rm D}_{[1} {\rm D}_{2]}) \bs K = ({\rm D}_{[1} {\rm D}_{3]} + {\rm D}_{[2} {\rm D}_{4]}) \bs L \\
        \label{e:CLSprimaryImaginary}
        {\rm D}_a {\rm D}_b  ((\gamma^5)^{ab} \bs K + i C^{ab} \bs L) &= 0 ~~~\Rightarrow ~~~ ({\rm D}_{[1} {\rm D}_{3]} + {\rm D}_{[2} {\rm D}_{4]})\bs K = - ({\rm D}_{[3} {\rm D}_{4]} - {\rm D}_{[1} {\rm D}_{2]})\bs L
  \end{align}
\end{subequations}
which result in the following simple constraints:
  \begin{subequations}\label{e:MNsubs}
  \begin{align}
       \label{e:Msubs}
      \bs \phi_1 \equiv \bs M - \Tw{\bs M} = 0 \Rightarrow  \bs Y_{13} + \bs Y_{24} = & \bs X_{34} - \bs X_{12}  \\
      \label{e:Nsubs}
      \bs \phi_2 \equiv \bs N - \Tw{\bs N} = 0  \Rightarrow  \bs X_{13} + \bs X_{24} = &\bs Y_{12} - \bs Y_{34}.
  \end{align}
  \end{subequations}
  After calculating either of 
  \begin{equation}\label{e:DaMN}
      {\rm D}_a \bs \phi_1 = 0~~~\mbox{or}~~~{\rm D}_a \bs \phi_2 =0
\end{equation}   
we uncover the following constraints on the fermions 
  \begin{equation}\label{e:LLt}
  \bs \Psi_a \equiv {\bs \Lambda}_a - \Tw{\bs \Lambda}_a - (\gamma^\mu)_a^{~d}\partial_\mu (\bs \zeta_d - \bs \rho_d) = 0.
  \end{equation}
  Defining
  \begin{equation}
     2{\bs \beta}_a \equiv   {\bs \Lambda}_a  + \Tw{\bs \Lambda}_a - (\gamma^\mu)_{a}^{~b}\partial_\mu ( \bs \zeta_b +  \bs \rho_b )
  \end{equation}
  we next eliminate ${\bs \Lambda}_a$ and $\Tw{\bs \Lambda}_a$ from the system by solving for them in terms of $\Tw{\bs \beta}_a$ 
\begin{subequations}
  \label{e:Lambdasubs}
  \begin{alignat}9
    {\bs \Lambda}_a = & {\bs \beta}_a + (\gamma^\mu)_a^{~d}\partial_\mu \bs \zeta_d  \label{e:Lambdasub}\\
    \Tw{\bs \Lambda}_a = & {\bs \beta}_a +  (\gamma^\mu)_a^{~d}\partial_\mu \bs \rho_d ~~~.\label{e:Lambdahatsub}
  \end{alignat}
\end{subequations}
  Our final two constraints are
  \begin{subequations}\label{e:DDMN}
\begin{align}\label{e:DDMNdtilde}
  (\gamma^5)^{ab} {\rm D}_a{\rm D}_b \bs \phi_1 = 0~~~&\mbox{or}~~~C^{ab}{\rm D}_a{\rm D}_b \bs \phi_2 = 0~~~\Rightarrow ~~~ \bs \Phi_1 \equiv \Tw{\bs {\rm d}} + 2 \partial_\mu \bs U^\mu + \square \bs L =0\\
  \label{e:DDMNd}
  C^{ab} {\rm D}_a{\rm D}_b \bs \phi_1 = 0~~~&\mbox{or}~~~(\gamma^5)^{ab}{\rm D}_a{\rm D}_b \bs \phi_2 = 0~~~\Rightarrow ~~~  \bs \Phi_2 \equiv {\bs {\rm d}} - 2 \partial_\mu \bs V^\mu  + \square \bs K = 0 
\end{align}  
\end{subequations}
      with $\square \equiv \partial_\mu \partial^\mu = \eta^{\mu\nu}\partial_\mu \partial_\nu $.
      This leaves a supermultiplet parametrized by the remaining $(2|8|10|4|0)$ component fields:
\begin{equation}
  \big(\,\bs K,\bs L\,\mid\,\bs \zeta_a,\bs \rho_b
        \mid \bs M, \bs N, \bs U_\mu, \bs V_\nu
         \mid \bs \beta_a \big),
 \label{e:CLSc0}
\end{equation}
\noindent Owing to the component field substitutions \eq{e:Lambdasubs} and \eq{e:DDMN}, these $12\,{+}\,12$ component fields indeed span the complete supermultiplet, reduced from the initial $16\,{+}\,16$ component fields by means of the constraints\eq{e:MNsubs}. The transformation laws of the resulting 12 + 12 component complex linear superfield are as in Eq.~(\ref{e:PreB}).

  \begin{subequations}
\label{e:PreB}
\begin{align}
 \rD_a\,\bs K &= \bs\z_a \\[1mm]
 \rD_a\,\bs L &= i(\g^5)_a^{~b} \bs\ro_b \\[2mm]
 \rD_a\,\bs\ro_b &= i C_{ab}\,\bs M  + (\g^5)_{a b}\,\bs N
                +i(\g^\m)_{ab}\,\bs V\!_\m - (\g^5 \g^\m)_{ab}\partial_\m \bs L \\[1mm]
 \rD_a\,\bs\z_b &= i C_{ab}\,\bs M  + (\g^5)_{a b}\,\bs N 
              +( \g^5 \g^\mu)_{ab}\,\bs U_\mu +  i(\g^\m)_{ab}\,\vd_\m\bs K \\[2mm]
 \rD_a\,\bs M &= \frc{1}2\bs\b_a
                \\[0mm]
 \rD_a\,\bs N &=  -i \frc{1}2 (\g^5)_a{}^b\,\bs\b_b
                  \\[0mm]
 \rD_a\,\bs U_\mu &= i\frc{1}2(\g^5 \g_\mu)_a{}^b\,\bs\b_b - i\frc{1}2(\g^5 \left[ \g^\n , \g_\m \right])_a{}^b\,\vd_\n\bs \zeta_b
                    \\[1mm]
 \rD_a\,\bs V_\mu &= - \frc{1}2(\g_\mu)_a{}^b\,\bs\b_b + \frc{1}2(\left[ \g^\n , \g_\m\right] )_a{}^b\,\vd_\n \bs \ro_b \\[2mm]
 \rD_a\,\bs\b_b &= 2 i (\g^\m)_{ab}\,\vd_\m\bs M
                  + 2 (\g^5\g^\m)_{ab}\,\vd_\m\bs N 
                  + 2(\g^5)_{ab}\,\vd^\m\bs U_\m +2 i C_{ab}\,\vd^\m\bs V_\m ~~~ .
\end{align}
\end{subequations}

Noting that the redefinitions \eq{e:Lambdasubs} and \eq{e:DDMN} indicate
 ({\small\bf1})~``1--several'' component field replacements and
 ({\small\bf2})~include second derivatives, the reader might suspect the latter of these to imply a violation of the Haag-{\Lv}opusa\'nski-Sohnius theorem in that a supercharge might transform a component field into another one the engineering dimension of which differs from the original one by more than $\pm\frc12$. In fact, the (standard\cite{r1001,rBK} and) manifestly supersymmetric complex superfield formulation and treatment of \CLS\ reassures us that this is not the case, and the appearance of the second derivatives merely points to the inadequacy of the basis\eq{e:CLSc0}.

 In turn, we show that the former of these features implies that the \CLS\ is not adinkraic; this result may simply herald the ``inconvenient truth:'' 
 that most\ft{As the ``1--several'' component field replacements\eq{e:Lambdasubs} and  \eq{e:DDMN} stem from the {\em\/simplicity\/} of the superdifferential constraints\eq{e:MNsubs} and the somewhat obvious fact that there exists indefinitely more non-{\em\/simple\/} superdifferential constraints\cite{rHSS}, the subset of {\em\/simple\/} superdifferential constraints and so-defined {\em\/simple\/} supermultiplets is of measure zero in the totality of all superdifferential constraints and so-defined supermultiplets.} representations of supersymmetry are not adinkraic. This agrees with the recent result of Ref.\cite{rTHGK12}.
 
\begin{figure}[htb]
% \begin{center}  % for if we want to use the .png's
% \begin{picture}(160,68)(5,0)%picture environment to input a .pdf to tag over or to put the cropped .png's in
% \put(0,0){\includegraphics[scale = 0.15]{Pix/BatwingLabelledRSSRPS.png}}
% \put(80,12.5){\includegraphics[scale=0.1]{Pix/BatwingLabelledArrow.png}}
% \put(98,.5){\includegraphics[scale = 0.15]{Pix/BatwingLabelledCLS.png}}
  \begin{center}
 \setlength{\unitlength}{.95mm} % this changes the units throughout, but within this "center" environment
 \begin{picture}(160,68)(10,0)
\put(0,0){\includegraphics[width=0.95\columnwidth]{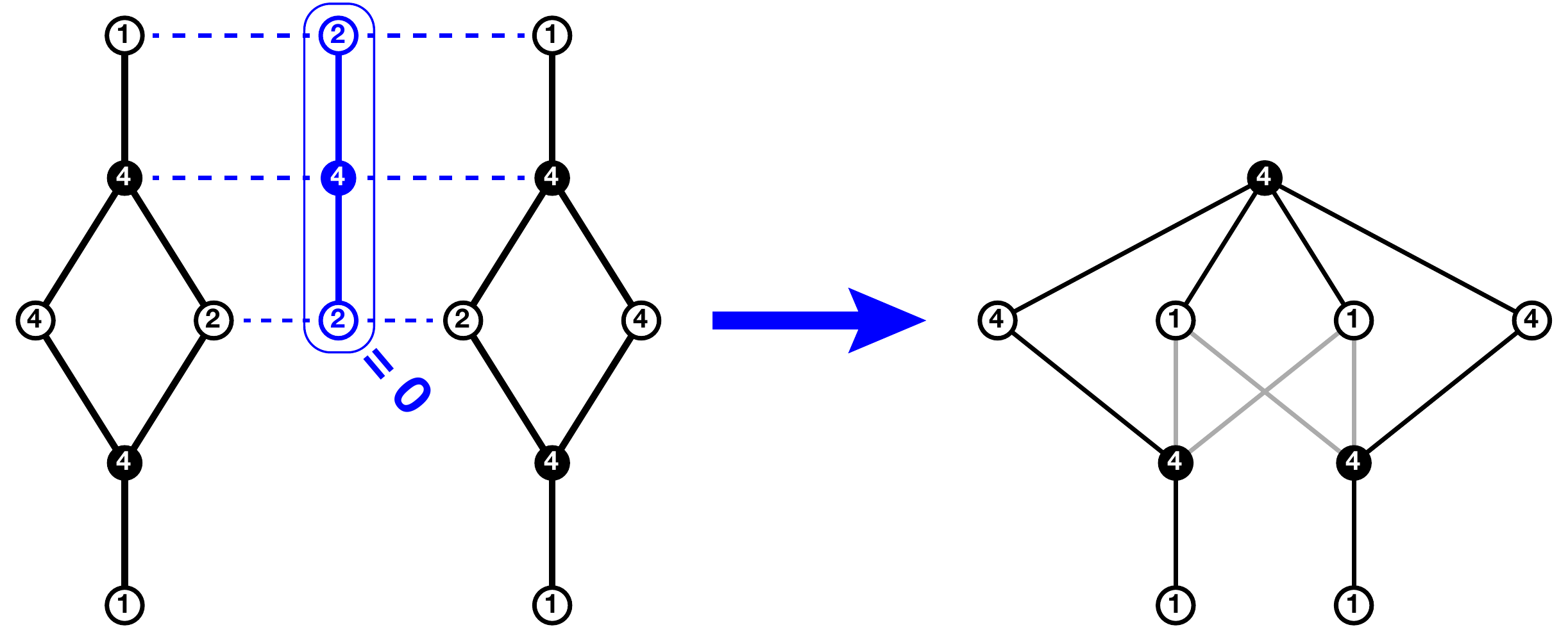}} % rescales the image
     %
     %RSS
     %
     \put(14.15,67.3){\cB{\hspace{1 pt}$\bs {\rm d} $\hspace{1 pt}}}
     \put(14.15,51.5){\CB{$\bs \Lambda_a$}}
     \put(4,35.7){\cB{$\bs U_\mu $}}
     \put(23,35.7){\cB{$\bs M , \bs N$}}     
     \put(14.15,19.9){\CB{$\bs \zeta_a$}}
     \put(14.15,2.5){\cB{$\bs K$}}    
     %
     %
     %Blue diagram
     %     
     \put(38.5,67.3){\cB{$\bs \Phi_i$}}
     \put(38.5,51.5){\cB{$\bs \Psi_a$}}
     \put(38.5,35.7){\cB{$\bs \phi_i$}}
     %
     %
     %RPS
     %
     \put(62.65,66.5){\cB{\hspace{1 pt}$\Tw{\bs {\rm d}}$\hspace{1 pt}}}
     \put(62.65,50.8){\CB{$\Tw{\bs \Lambda}_a$}}
     \put(73,35.7){\cB{$\bs V_\mu $}}
     \put(54,35){\cB{$\Tw{\bs M} , \Tw{\bs N}$}}     
     \put(62.65,19.9){\CB{$\bs \rho_a$}}
     \put(62.65,2.5){\cB{$\bs L$}}    
     %
     %
     %CLS
     %
     \put(143.5,51.5){\CB{$\bs \beta_a$}}
     \put(114,35){\cB{$\bs U_\mu $}}
     \put(133,34.9){\cB{$\bs M$}} 
     \put(153.7,34.9){\cB{$\bs N$}}      
     \put(172.5,35){\cB{$\bs V_\mu $}}    
     \put(133.5,19.3){\CB{$\bs \zeta_a$}}
     \put(153.5,19.9){\CB{$\bs \rho_a$}}
     \put(133.5,2.5){\cB{$\bs K$}}    
     \put(153.5,2.5){\cB{$\bs L$}}        
 \end{picture}
 \end{center}
 \caption{A schematic of the `zippering' process that fuses the \RSS\ and \RPS\ (left) into the \CLS\ (right). The black edges indicate bi-directional transformations and the grey edges indicate uni-directional transformations. For instance, the black edge $\bs M$\protect\rule[.5ex]{10pt}{1pt}$\bs \beta_a$ indicates that $ \rD_a M \supset \bs \beta_a $ and $ \rD_a \bs \beta_b \supset \bs M $, where as the grey edge $\bs M$\textcolor{grey3}{\protect\rule[.5ex]{10pt}{1pt}}$\bs \zeta_a$ indicates that, e.g., $\mbox{\textcolor{grey3}{$\rD_a$}} \bs \zeta_b \supset \bs M $ but $\mbox{\textcolor{grey3}{$\rD_a$}} \bs M \not\supset \bs \zeta_a$. 
 }
 \label{f:Batwng}
\end{figure}

 As was noted in~\cite{r6-4.2}, the constraints in~\eq{e:MNsubs},~\eq{e:LLt}, and~\eq{e:DDMN} 
 \begin{equation}
 \big(\, \bs \phi_1 , \, \bs \phi_2 \,\mid\, \bs \Psi_a \, \mid \, \bs \Phi_1 , \, \bs \Phi_2 \, \big),
\label{e:ChRLcnstrnt}
\end{equation}
\noindent where
\begin{subequations}
\begin{align}
      \bs \phi_1 \equiv & \bs M - \Tw{\bs M}~~~,~~~\bs \phi_2 \equiv \bs N - \Tw{\bs N}~~~, \\
      \bs \Psi_a \equiv & {\bs \Lambda}_a - \Tw{\bs \Lambda}_a - (\gamma^\mu)_a^{~d}\partial_\mu (\bs \zeta_d - \bs \rho_d) \\
      \bs \Phi_1 \equiv & \Tw{\bs {\rm d}} + 2 \partial_\mu \bs U^\mu + \square \bs L~~~,~~~\bs \Phi_2 \equiv {\bs {\rm d}} - 2 \partial_\mu \bs V^\mu  + \square \bs K ~~~,
\end{align}
\end{subequations}
\noindent 
 themselves form a chiral supermultiplet.  Thus, if we draw the Adinkras for the \RSS\ 
 and the \RPS\ the imposition of the constraints can be indicated by the insertion of the 
 Adinkra of a chiral multiplet relating to the components of the \RSS\ and the \RPS\  
 which is then set to zero.  This chiral multiplet is illustrated in the `blue region' of Fig.~\ref{f:Batwng}
 \noindent 
where the rightmost Adinkra corresponds to the fields and transformation laws in
Eq.~\eq{e:PreB}.  Figure~\ref{f:Batwng} is a `blueprint' for the \CLS\ Adinkra with explicit field content which we will build in Section~\ref{s:CLS0red}. The process depicted in Fig.~\ref{f:Batwng} was called `zippering' in the work of Ref.~\cite{rDFGHIL08}.  However, K.~Iga has noted that the process imposed on the Adinkra in someways resembles the process of constructing a gnomon~\cite{rGnomon} in classical geometry.
  
  \subsection{0-brane Reduction and Adinkra}\label{s:CLS0red}
The two constraints~\eq{e:MNsubs} allow us to remove $X_{13}$ and $Y_{13}$ from the resulting dimensionally reduced complex linear supermultiplet, leaving us with the linearly independent set:

\begin{equation}
  \big(\,\bs K,\bs L\,\mid\,\bs \zeta_I,\bs \rho_J \,\mid\,
        \bs X_{12}, \bs X_{14}, \bs X_{23}, \bs X_{24}, \bs X_{34}, \bs Y_{12}, \bs Y_{14}, \bs Y_{23}, \bs Y_{24}, \bs Y_{34}
         \mid \bs \beta_I \big).
 \label{e:CLSXYM}
\end{equation}
where
\begin{align}\label{e:CLSXY}
    \bs X_{12} = &  \bs U_2 - \bs M,~~~\bs  X_{23} = \bs U_1 - \bs U_0, ~~~\bs X_{24} = -\bs U_3 - \bs N, \cr
 &\bs X_{14} =  \bs U_0 + \bs U_1,~~~\bs X_{34} = \bs U_2 + \bs M, \cr
  \bs Y_{12} = & \bs V_2 -  \bs N,~~~ \bs Y_{23} =  \bs V_1 - \bs V_0,~~~ \bs Y_{24} = \bs M - \bs V_3 \cr
  & \bs Y_{14} =  \bs V_0 + \bs V_1,~~~ \bs Y_{34} = \bs V_2 + \bs N
\end{align}
The dimensionally reduced versions of the substitutions~\eq{e:Lambdasubs} and~\eq{e:DDMN} which take us from \RSS\ and \RPS\ union to \CLS\ are
\begin{subequations}\label{e:0subs}
\begin{align}\label{e:LambdasubsExpanded}
    \bs \Lambda_1 =&  \bs \beta_1 + \Dt{\bs \zeta}_2~~~,~~~\bs \Lambda_2 =  \bs \beta_2 {-} \Dt{\bs \zeta}_1 ~~~,~~~    \bs \Lambda_3 =  \bs \beta_3 {-} \Dt{\bs \zeta}_4 ~~~,~~~\bs \Lambda_4 =  \bs \beta_4 {+} \Dt{\bs \zeta}_3 ~~~, \\
    \label{e:tLambdasubsExpanded}
    \Tw{\bs \Lambda}_1 =&  \bs \beta_1 + \Dt{\bs \rho}_2 ~~~,~~~\Tw{\bs \Lambda}_2 =  \bs \beta_2 {-} \Dt{\bs \rho}_1 ~~~,~~~
    \Tw{\bs \Lambda}_3 =  \bs \beta_3 {-} \Dt{\bs \rho}_4 ~~~,~~~\Tw{\bs \Lambda}_4 =  \bs \beta_4 {+} \Dt{\bs \rho}_3 ~~~, \\
    \label{e:dsubsExpanded}
    \bs {\rm d} =&  - 2 \Dt{\bs V_0} + \DDt{\bs K} = \Dt{\bs Y}_{23} - \Dt{\bs Y}_{14} + \DDt{\bs K}~~~,~~~\Tw{\bs {\rm d}} = 2\Dt{\bs U_0} + \DDt{\bs L} = \Dt{\bs X}_{14} - \Dt{\bs X}_{23} + \DDt{\bs L}
\end{align}
\end{subequations}

We now copy the dimensionally reduced transformation laws in Tables~\ref{t:ScX}and~\ref{t:ScY} for the real scalar and pseudoscalar supermultiplets, respectively (utilizing the substitutions \eq{substitution1expanded}), and perform the substitutions~\eq{e:MNsubs},~\eq{e:CLSXY}, and~\eq{e:0subs}, omitting also the substituted fields, $\bs X_{13} ,\bs Y_{13}, \bs \Lambda_I,\Tw{\bs \Lambda}_J, \bs {\rm d}$ and $\Tw{\bs {\rm d}}$, from the so-obtained Table~\ref{t:ScXY}.
\begin{table}[htb]
\caption{The dimensionally reduced complex linear supermultiplet supersymmetry transformation rules: the initial basis\eq{e:CLSc0}.}
\vspace*{-2mm}\small
$$\begin{array}{@{} c|cccc @{}}
 \omit& \C3{\rD_1}  & \C1{\rD_2} & \C6{\rD_3} & \C2{\rD_4} \strut\\ 
    \toprule
 \bs K & \bs \zeta_1 & \bs \zeta_2 & \bs \zeta_3 & \bs \zeta_4 \\
 \bs L & -\bs \rho_4 & \bs \rho_3 & -\bs \rho_2 & \bs \rho_1 \\
 \midrule
 \bs \zeta_1   & i\Dt{\bs K} & -i\bX_{12} & i(\bs X_{24} - \bs Y_{12} + \bs Y_{34}) & -i\bX_{14}\\
 \bs \zeta_2   & i\bX_{12} & i\Dt{\bs K}  & -i\bX_{23} & -i\bX_{24}\\
 \bs \zeta_3  & -i(\bs X_{24} - \bs Y_{12} + \bs Y_{34}) & i\bX_{23}  & i\Dt{\bs K}  & -i \bs X_{34}\\
 \bs \zeta_4   & i \bX_{14} & i \bX_{24}  & i \bs X_{34}  & i \Dt{\bs K}\\
 \bs \rho_1   & i \bs Y_{14} & i \bs Y_{24} & i \bs Y_{34} & i\Dt{\bs L}\\
 \bs \rho_2   & i(\bs Y_{24} + \bs X_{12} - \bs X_{34}) & -i\bs Y_{23}  & -i\Dt{\bs L} & i \bs Y_{34}\\
 \bs \rho_3   & i \bs Y_{12} & i\Dt{\bs L}  & -i \bs Y_{23}  & -i \bs Y_{24}\\
 \bs \rho_4   & -i\Dt{\bs L} & i \bs Y_{12}  & -i(\bs Y_{24} + \bs X_{12} - \bs X_{34}) & i \bs Y_{14}\\
 \midrule
 \bX_{12} & \Dt{\bs \zeta}_2 & -\Dt{\bs \zeta}_1 & - \bs \beta_3 + \Dt{\bs \zeta}_4 
                                   & - \bs \beta_4 -  \Dt{\bs \zeta}_3  \\
 \bX_{14} & \Dt{\bs \zeta}_4 &  \bs \beta_4 + \Dt{\bs \zeta}_3    & - \bs \beta_1 - \Dt{\bs \zeta}_2  & -\bs \zeta_1\\
 \bX_{23} & -  \bs \beta_3 + \Dt{\bs \zeta}_4   & \Dt{\bs \zeta}_3    &  -\Dt{\bs \zeta}_2&  \bs \beta_2 - \Dt{\bs \zeta}_1 \\
 \bX_{24} & - \bs \beta_4 - \Dt{\bs \zeta}_3    & \Dt{\bs \zeta}_4  & - \bs \beta_2 + \Dt{\bs \zeta}_1  & -\Dt{\bs \zeta}_2\\
 \bX_{34} &  \bs \beta_1 + \Dt{\bs \zeta}_2 & \bs \beta_2 - \Dt{\bs \zeta}_1   & \Dt{\bs \zeta}_4  & -\Dt{\bs \zeta}_3\\
 \bY_{12} & \Dt{\bs \rho}_3 & \Dt{\bs \rho}_4 & {-} \bs \beta_2 {+} \Dt{\bs \rho}_1
                                   & \bs \beta_1 {+}  \Dt{\bs \rho}_2\\
 \bY_{14} & \Dt{\bs \rho}_1  & {-} \bs \beta_1 {-}  \Dt{\bs \rho}_2 & {-} \bs \beta_4 {-} \Dt{\bs \rho}_3 & \Dt{\bs \rho}_4\\
 \bY_{23} & {-} \bs \beta_2 {+} \Dt{\bs \rho}_1 & -\Dt{\bs \rho}_2   & -\Dt{\bs \rho}_3   & {-} \bs \beta_3 {+} \Dt{\bs \rho}_4\\
 \bY_{24} &  \bs \beta_1 {+} \Dt{\bs \rho}_2    & \Dt{\bs \rho}_1  &  \bs \beta_3 {-} \Dt{\bs \rho}_4 & -\Dt{\bs \rho}_3\\
 \bY_{34} &  \bs \beta_4 {+} \Dt{\bs \rho}_3   & {-} \bs \beta_3 {+} \Dt{\bs \rho}_4  & \Dt{\bs \rho}_1     & \Dt{\bs \rho}_2\\
 \midrule
 \bs \beta_1   & i ( \Dt{\bX}_{34} - \Dt{\bs X}_{12})  
         & -i (\Dt{\bs Y}_{14} {-} \Dt{\bs Y}_{23}) 
        & -i(\Dt{\bs X}_{14} {-} \Dt{\bs X}_{23}) 
        & -i(\Dt{\bs Y}_{34} {-} \Dt{\bs Y}_{12}  )
         \\
 \bs \beta_2   & i(\Dt{\bs Y}_{14} {-} \Dt{\bs Y}_{23} ) 
         & i(\Dt{\bs X}_{34} {-} \Dt{\bs X}_{12} ) 
        & i(\Dt{\bs Y}_{34} {-} \Dt{\bs Y}_{12} ) 
        & -i(\Dt{\bs X}_{14} {-} \Dt{\bs X}_{23} ) 
        \\
 \bs \beta_3  & i(\Dt{\bs X}_{14} {-} \Dt{\bs X}_{23}  ) 
         & -i(\Dt{\bs Y}_{34} {-} \Dt{\bs Y}_{12}) 
        & i(\Dt{\bs X}_{34} {-} \Dt{\bs X}_{12})  & 
        i(\Dt{\bs Y}_{14} {-} \Dt{\bs Y}_{23} ) \\
 \bs \beta_4   & i(\Dt{\bs Y}_{34} {-} \Dt{\bs Y}_{12}) 
         & (\Dt{\bs X}_{14} {-} \Dt{\bs X}_{23}  ) 
        & -i(\Dt{\bs Y}_{14} {-} \Dt{\bs Y}_{23} )  
        & i(\Dt{\bs X}_{34} {-} \Dt{\bs X}_{12}) \\
    \bottomrule
  \end{array}$$
\label{t:ScXY}
\end{table}

\paragraph{Higher-Level Bosons:}
The transformation results of $\rD_I\,\bs \zeta_J$ and $\rD_I\,\bs \rho_J$ involve 6+6 distinct linear combinations of the 5+5 independent bosons $\bX_{IJ}$ and $\bY_{IJ}$. It is therefore impossible to rewrite each of these as a single, linearly independent component field or its derivative, and the \CLS\ supermultiplet is therefore unavoidably non-adinkraic.

We thus turn to the ten higher-level bosons, $\bX_{IJ},\bY_{IJ}$, with $IJ\neq13$, and notice that the particular combinations $\bX_{34}{-}\bX_{12}$ and $\bY_{34}{-}\bY_{12}$ occur
 ({\small\bf1})~throughout the supersymmetry transformations of
       $\bs \beta_I$, as well as
 ({\small\bf2})~in {\em\/all\/} trinomial terms in the supersymmetry
       transformations of $\bs \zeta_1,\bs \zeta_3,\bs \rho_2,\bs \rho_4$.
Therefore, the redefinitions
\begin{subequations}
 \label{e:HBr}
\begin{equation}
 \bX_{34}\to\eX_{34}\Defl(\bX_{34}{-}\bX_{12}) \quad\text{and}\quad
 \bY_{34}\to\eY_{34}\Defl(\bY_{34}{-}\bY_{12})
 \label{e:HBrA}
\end{equation}
reduce all trinomials in the results of $\rD_I\,\bs \zeta_J$ and $\rD_I\,\bs \rho_J$ into binomials, and also reduce some of the binomials in the results of $\rD_I\,\bX_{JK}$ and $\rD_I\,\bY_{JK}$.
 Similar simplification are achieved also by the replacing
\begin{equation}
  \bX_{14}\to\eX_{14}\Defl(\bX_{14}{-}\bX_{23}) \quad\text{and}\quad
  \bY_{14}\to\eY_{14}\Defl(\bY_{14}{-}\bY_{23}).
 \label{e:HBrB}
\end{equation}
\end{subequations}
The component field combinations\eq{e:HBrA} are related to the primary relations\eq{e:MNsubs}, while the component field combinations\eq{e:HBrB} are related to the secondary relation \eq{e:dsubsExpanded}.

\begin{figure}[htb]
 \begin{center}
  \begin{picture}(160,55)(0,14)
   \put(0,7){\includegraphics[width=160mm]{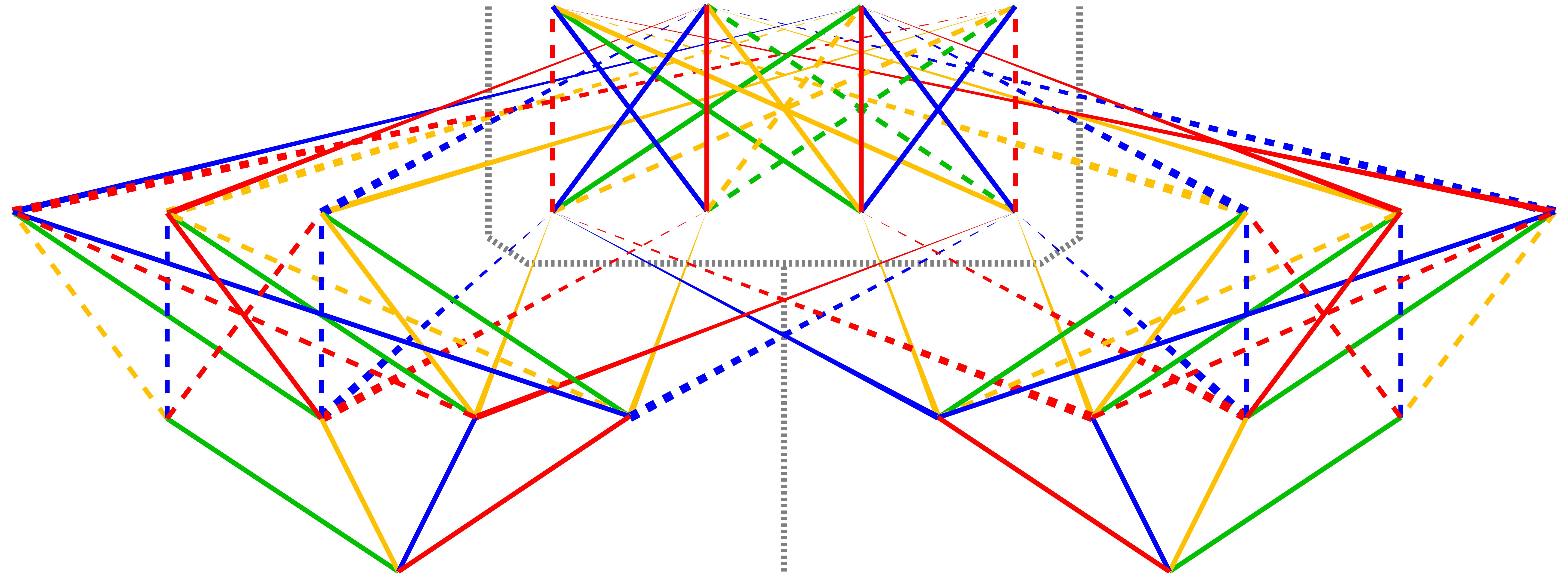}}
     \put(57,66){\CB{$\bs \beta_1$}}
     \put(72,66){\CB{$-\bs \beta_3$}}
     \put(88,66){\CB{$-\bs \beta_2$}}
     \put(103,66){\CB{$\bs \beta_4$}}
     \put(4,43){\cB{$\bX_{24}$}}
     \put(19,43){\cB{$\bX_{23}$}}
     \put(34,43){\cB{$\bX_{21}$}}
     \put(57,43){\cB{$\eX_{43}$}}
     \put(72,43){\cB{$\eX_{41}$}}
     \put(88,43){\cB{$\eY_{41}$}}
     \put(103,43){\cB{$\eY_{43}$}}
     \put(126,43){\cB{$\bY_{21}$}}
     \put(141,43){\cB{$\bY_{23}$}}
     \put(156,43){\cB{$\bY_{24}$}}
     \put(17,22.5){\CB{$\bs \zeta_2$}}
     \put(33,22.5){\CB{$\bs \zeta_4$}}
     \put(49,22.5){\CB{$\bs \zeta_3$}}
     \put(65,22.5){\CB{$\bs \zeta_1$}}
     \put(96,23){\CB{$-\bs \rho_4$}}
     \put(112,23){\CB{$-\bs \rho_2$}}
     \put(127,23){\CB{$\bs \rho_1$}}
     \put(143,23){\CB{$\bs \rho_3$}}
     \put(40.5,7){\cB{$\bs K$}}
     \put(119.5,7){\cB{$\bs L$}}
  \end{picture}
 \end{center}
 \caption{A quasi-Adinkraic graphical depiction of the \CLS. }
 \label{f:CLS1}
\end{figure}
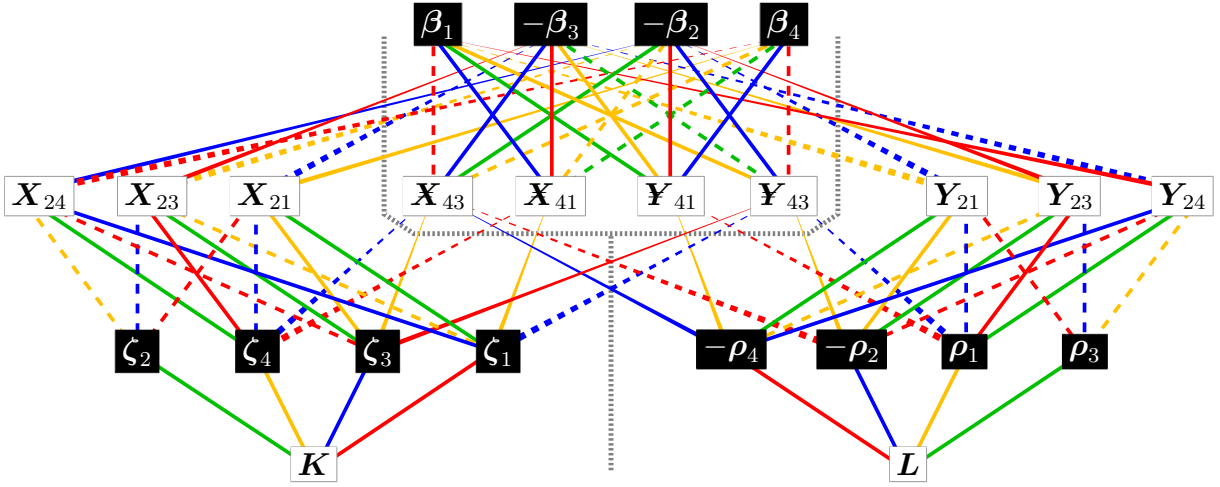

\begin{table}[htb]
\caption{The superderivative relations for the dimensionally reduced complex linear superfield, after the field redefinitions~\eq{e:HBr}. The fields and colors are organized to expose the antisymmetric $F_{\mu\nu}$-like structure in the transformations.}
\vspace*{-2mm}\footnotesize
$$\begin{array}{@{} c|cccc @{}}
 & \C1{\rD_2}  & \C2{\rD_4} & \C6{\rD_3} & \C3{\rD_1}\strut\\ 
    \toprule\vspace*{-2pt}
 \bs K & \bs \zeta_2 & \bs \zeta_4 & \bs \zeta_3 & \bs \zeta_1 \\
 \bs L & \bs \rho_3 & \bs \rho_1 & -\bs \rho_2 & -\bs \rho_4\\
 \midrule
 \bX_{24} & \Dt{\bs \zeta}_4 & -\Dt{\bs \zeta}_2 & \Dt{\bs \zeta}_1\Gr{{-}\bs \beta_2} &-\Dt{\bs \zeta}_3\Gr{{-}\bs \beta_4}\\
 \bX_{23} & \Dt{\bs \zeta}_3 & -\Dt{\bs \zeta}_1\Gr{{+}\bs \beta_2}& -\Dt{\bs \zeta}_2 & \Dt{\bs \zeta}_4\Gr{{-}\bs \beta_3}\\
 \bX_{21} & \Dt{\bs \zeta}_1 & \Dt{\bs \zeta}_3\Gr{{+}\bs \beta_4} & -\Dt{\bs \zeta}_4\Gr{{+}\bs \beta_3} & -\Dt{\bs \zeta}_2\\[1pt]
 \bY_{24} & \Dt{\bs \rho}_1 & -\Dt{\bs \rho}_3 & -\Dt{\bs \rho}_4\Gr{{+}\bs \beta_3} & \Dt{\bs \rho}_2\Gr{{+}\bs \beta_1}\\
 \bY_{23} & -\Dt{\bs \rho}_2 & \Dt{\bs \rho}_4\Gr{{-}\bs \beta_3} & -\Dt{\bs \rho}_3 & \Dt{\bs \rho}_1\Gr{{-}\bs \beta_2}\\
 \bY_{21} & -\Dt{\bs \rho}_4 & -\Dt{\bs \rho}_2\Gr{{-}\bs \beta_1} & -\Dt{\bs \rho}_1\Gr{{+}\bs \beta_2} & -\Dt{\bs \rho}_3\\[1pt]
  \cline{2-5}\rule{0pt}{3.3ex}
 \Gr{\eX_{43}} & \Gr{-\bs \beta_2}  & \Gr{-\bs \beta_4} & \Gr{-\bs \beta_3} & \Gr{-\bs \beta_1}\\
 \Gr{\eX_{14}} & \Gr{\bs \beta_4} & \Gr{-\bs \beta_2} & \Gr{-\bs \beta_1} & \Gr{\bs \beta_3}\\
 \Gr{\eY_{43}} & \Gr{\bs \beta_3} & \Gr{\bs \beta_1}  & \Gr{-\bs \beta_2} & \Gr{-\bs \beta_4}\\
 \Gr{\eY_{41}} & \Gr{\bs \beta_1}  & \Gr{-\bs \beta_3}  & \Gr{\bs \beta_4} & \Gr{-\bs \beta_2}\\[-1pt]
    \bottomrule
 \end{array}\qquad
\begin{array}{@{} c|cccc @{}}
  & \C1{\rD_2}  & \C2{\rD_4} & \C6{\rD_3} & \C3{\rD_1} \strut\\ 
    \toprule
 \bs \zeta_2   & i\Dt{\bs K} & -i \bX_{24} & -i \bX_{23} & -i\bX_{21}\\[1mm]
 \bs \zeta_4   & i\bX_{24} & i\Dt{\bs K}  & -i\bX_{21}\Gr{{-i}\eX_{43}} & i\bX_{23}\Gr{{-i}\eX_{41}}\\[1pt]
 \bs \zeta_3   & i\bX_{23} & i\bX_{21}\Gr{{+i}\eX_{43}} & i\Dt{\bs K}  & -i\bX_{24}\Gr{{+i}\eY_{43}}\\[1pt]
 \bs \zeta_1   & i\bX_{21} & -i\bX_{23}\Gr{{+i}\eX_{41}} & i\bX_{24}\Gr{{-i}\eY_{43}}  & i\Dt{\bs K}\\[1pt]
 \bs \rho_3   & i\Dt{\bs L} & -i\bY_{24} & -i\bY_{23} & -i\bY_{21}\\[1mm]
 \bs \rho_1   & i\bY_{24} & i\Dt{\bs L}  & -i\bY_{21}\Gr{{-i}\eY_{43}} & i\bY_{23}\Gr{{-i}\eY_{41}}\\[1pt]
 -\bs \rho_2   & i\bY_{23} & i\bY_{21}\Gr{{+i}\eY_{43}} & i\Dt{\bs L}  & -i\bY_{24}\Gr{{-i}\eX_{43}}\\[1pt]
 -\bs \rho_4   & i\bY_{21} & -i\bY_{23}\Gr{{+i}\eY_{41}} & i\bY_{24}\Gr{{+i}\eX_{43}} & i\Dt{\bs L}\\[1mm]
 \midrule
 \Gr{\bs \beta_1} &  i\Gr{\Dt\eY_{41}} & i\Gr{\Dt\eY_{43}} &  i\Gr{\Dt\eX_{41}} & -i\Gr{\Dt\eX_{43}}\\[1pt]
 \Gr{-\bs \beta_3} & -i\Gr{\Dt\eY_{43}} &  i\Gr{\Dt\eY_{41}} &  i\Gr{\Dt\eX_{43}} &  i\Gr{\Dt\eX_{41}}\\[1pt]
 \Gr{\bs \beta_4} & -i\Gr{\Dt\eX_{41}} & -i\Gr{\Dt\eX_{43}} &  i\Gr{\Dt\eY_{41}} & -i\Gr{\Dt\eY_{43}}\\[1pt]
 \Gr{-\bs \beta_2} &  i\Gr{\Dt\eX_{43}} & -i\Gr{\Dt\eX_{41}} &  i\Gr{\Dt\eY_{43}} &  i\Gr{\Dt\eY_{41}}\\
    \bottomrule
  \end{array}$$
\label{t:ScXY++}
\end{table}%
These replacements turn Table~\ref{t:ScXY}  into Table~\ref{t:ScXY++}, written in terms of the optimal basis~\eq{e:CLSXYMslash}
\begin{equation}
  \big(\,\bs K,\bs L\,\mid\,\bs \zeta_I,\bs \rho_J \,\mid\,
        \bs X_{21}, \bs X_{23}, \bs X_{24}, \bs Y_{21}, \bs Y_{23}, \bs Y_{24},  \eX_{41}, \eX_{43}, \eY_{41}, \eY_{43}
         \mid \bs \beta_I \big).
 \label{e:CLSXYMslash}
\end{equation}
 The structure shown in Table~\ref{t:ScXY++} may be depicted akin to the Adinkras of Refs.\cite{r6-1,r6-3,r6-3.2,r6-1.2}, provided we introduce an additional graphical element: uni-directional edges, to represent the fact that, \eg, $\C2{\rD_4}\,\bs \zeta_3\supset \eX_{43}$, but $\C2{\rD_4}\,\eX_{43}\not\supset\Dt{\bs \zeta}_3$.  (Nevertheless, as the results in Table~\ref{t:ScXY++} easily show, the algebra\eq{e:RDuSy} is satisfied by the $Q_I$-action upon each and every component field.)
 These uni-directional (partial) transformations were depicted as grey edges in the `blueprint' of Fig.~\ref{f:Batwng}. Uni-directional edges were also introduced in Ref.\cite{r6-4.2} where they too were depicted as grey edges. Herein, to avoid complicating the graphs using four additional colors and to visually indicate their (exclusively) upward direction, we depict uni-directional edges by tapering edges; see Figure~\ref{f:CLS1}.

 \paragraph{Top-Level Fermions:}
  Further row-operations in the same vein as those used for the higher-level bosons such as $\bs \beta_1\to(\bs \beta_1+\Dt{\bs \rho}_2)$ so as to remove the non-Adinkraicity from the transformations of the $\bs X_{IJ},\bs Y_{IJ}$ bosons would re-introduce second derivatives in the supersymmetry transformation rule:
\begin{equation}
  \C6{\rD_3}(\bs \beta_1+\Dt{\bs \rho}_2)
   = -i\Dt\eX_{14}-i\DDt{\bs L\,}.
% \label{e:}
\end{equation}
In turn, the lower-level fermions, $\bs \zeta_I,\bs \rho_I$ cannot be redefined by an admixture of the top-level fermions {\em\/locally\/}.
 Furthermore, no row-operation amongst the top-level fermions offers any cancellation or qualitative simplification. This leaves $\bs \beta_I$ as the optimal choice.

\paragraph{Reduction:}
Table~\ref{t:ScXY++} and Figure~\ref{f:CLS1} make it clear that the top-most component fields,
\begin{equation}
  \CLS_V \Defl
  (\eX_{43},\eX_{41},\eY_{41},\eY_{43}\mid\bs \beta_I),
 \label{e:CLSv}
\end{equation}
all by themselves properly close under supersymmetry, and so span a sub-representation, {\em\/within\/} \CLS. Furthermore, the sub-supermultiplet\eq{e:CLSv} is adinkraic and has the {\em\/chromotopography\/}\ft{Extending the definition of chromotopology\cite{r6-3}, {\em\/chromotopography\/} denotes ``chromotopology\,+\,height,'' \ie, the chromotopology with the nodes drawn at the height proportional to the engineering dimension of the corresponding component fields.} of a trans-{\em\/valise\/}~\cite{rUMD09-1}. That is to say, it is possible (in principle) to consistently gauge away the remaining lower components, $(\bs K,\bs L \mid\bs \zeta_I,\bs \rho_I\mid \bX_{2J},\bY_{2J})$, leaving behind the supermultiplet\eq{e:CLSv} as an analogue of the Wess-Zumino gauge-fixed ``vector'' supermultiplet\cite{r1001,rBK}.
Thereby, the \CLS\ is {\em\/reduced\/}, \ie, shown to contain smaller supermultiplets,  although it did not {\em\/decompose\/} into their direct sum.

Now, because of:
\begin{enumerate}\itemsep=-3pt\vspace{-2mm}
 \item the previous conclusion that the top-level fermions, $\bs \beta_I$ must not be mixed with (the $\vdt$-derivatives of) the lower-level fermions, and
 \item the fact that no row-operation in the rows with the $Q$-transformations of the component fields $\bX_{21},\bX_{23},\bX_{24},\bY_{21},\bY_{23},\bY_{24}$ can eliminate the remaining binomials,
\end{enumerate}
we conclude that the ``other'' component fields $(\bs K,\bs L \mid\bs \zeta_I,\bs \rho_I\mid \bX_{2J},\bY_{2J})$ do not form a closed, separate supermultiplet|unless we set
\begin{equation}
  (\eX_{41},\eX_{43},\eY_{41},\eY_{43}\mid\bs \beta_I)
   \Is (0,0,0,0|0,0,0,0),
% \label{e:}
\end{equation}
whereupon the remainder of \CLS, $\CLS|_{\CLS_V=0}$, clearly decomposes into
\begin{equation}
  \Big(\,\IT_x\Defl(\bs K \mid\bs \zeta_I\mid \bX_{21},\bX_{23},\bX_{24})\Big)
   ~\oplus~
  \Big(\,\IT_y\Defl(\bs L \mid\bs \rho_I\mid \bY_{21},\bY_{23},\bY_{24})\Big),
% \label{e:}
\end{equation}
which are also adinkraic, and are recognized as the worldline dimensional reduction of {\em\/real linear supermultiplets\/}, with dimensions $(1|4|3|0|0)$ each. Indeed, notice the $F_{\m\n}$-like structure of the $\rD_I$-transformations of $\bs \zeta_I$ and $\bs \rho_I$ in Table~\ref{t:ScXY++}.

Schematically, using only the dimensions of these adinkraic supermultiplets,
\begin{equation}
  \CLS = \Bm{0\\[1pt]4\\[1pt]10\\[1pt]8\\[1pt]2\\[1pt]} ~=~
   \left\{
     \mathop{\Bm{0\\[1pt]0\\[1pt]3\\[1pt]4\\[1pt]1\\[1pt]}}_{\IT_x}
       \raisebox{-2mm}{\rotatebox{30}{\shortstack[t]{$\,\to$\\[-1mm]$\nleftarrow\,$}}}
     \mathop{\Bm{0\\[1pt]4\\[1pt]4\\[1pt]0\\[1pt]0\\[1pt]}^{\strut}}_{\CLS_V}
       \raisebox{0mm}{\rotatebox{-30}{\shortstack[t]{$\from\,$\\[-1mm]$\,\nrightarrow$}}}
     \mathop{\Bm{0\\[1pt]0\\[1pt]3\\[1pt]4\\[1pt]1\\[1pt]}}_{\IT_y}
   \right\},
 \label{e:CLSr}
\end{equation}
where the arrows indicate that the two ``sideline'' supermultiplets $\IT_x$ and $\IT_y$ $\rD$- ($Q$-)transform into the ``central'' portion (with higher engineering dimension), $\CLS_V$, but not the other way around.

This is reminiscent of the ``kites'' in Ref.\cite[~Eq.\,(7)]{r6-4.2}, but the chromotopographies are rather different: here $\IT_x$ and $\IT_y$ are jointly flying the valise kite, $\CLS_V$. Owing to the ``bow-ties theorems'' of Ref.\cite{rGH-obs}, it is clear that this valise\eq{e:CLSr} by itself cannot  extend even to worldsheet supersymmetry (other than the trivial extension cases of unidextrous $(4,0)$- and $(0,4)$-supersymmetry), and therefore certainly does not extend to supersymmetry in any higher-dimensional spacetime. However, the ``bow-ties theorems" of Ref.\cite{rGH-obs} do not apply to gauge-fixed supermultiplets, and it is thus not impossible that $\CLS_V$ as a gauge-fixed supermultiplet ({\em\/\'a la\/} Wess-Zumino) does extend to higher dimensions. In fact, this observation brings us back
 to the very first presentation~\cite{rJGWS81} of the components of the complex linear multiplet.
 
   In this `ancient' work in Ref.~\cite{rJGWS81} there appears a discussion highlighting the fact that the fundamental superfield which describes the \CLS\ is actually a gauge spinor superfield.  This is true even though none of component fields in the multiplet are gauge fields.  Nonetheless the superfield that describes them is a gauge superfield.  In fact, the $\CLS_V$ submultiplet is the analog for the \CLS\ superfield as the Weyl supermultiplet is for the supergravity supermultiplet~\cite{rWSJG79}.

\paragraph{Encoding the Reduction:}
To specify the ``sideline'' ``tensor-like'' supermultiplets portions of \CLS\ as proper sub-super\-multi\-plets, we must impose a $Q$-covariant constraint on \CLS, one that would annihilate the $(0|0|4|4|0)$-dimensional ``central'' portion, $\CLS_V$, which holds the \CLS\ together. The $Q$-invariance of the superderivatives ($\{Q_I,\rD_J\}=0$) makes the manifestly supersymmetric superfield formalism the natural choice.

Thus, we wish to annihilate the bosons $\eX_{34}\Defl(\bX_{34}{-}\bX_{12})$, $\eX_{14}\Defl(\bX_{14}{-}\bX_{23})$, $\eY_{34}\Defl(\bY_{34}{-}\bY_{12})$, and $\eY_{14}\Defl(\bY_{14}{-}\bY_{23})$. These combinations are defined as the superderivatives
\begin{subequations}
% \label{e:}
\begin{alignat}9
 \eX_{34}&=-i\frc{1}2\big[\rD_{[3}\rD_{4]}-\rD_{[1}\rD_{2]}\big] \bs K,&\qquad
 \eY_{34}&=-i\frc{1}2\big[\rD_{[3}\rD_{4]}-\rD_{[1}\rD_{2]}\big]\bs L,\\
 \eX_{14}&=-i\frc{1}2\big[\rD_{[1}\rD_{4]}-\rD_{[2}\rD_{3]}\big]\bs K,&\qquad
 \eY_{14}&=-i\frc{1}2\big[\rD_{[1}\rD_{4]}-\rD_{[2}\rD_{3]}\big]\bs L.
% \label{e:}
\end{alignat}
\end{subequations}
This suggests the annihilation of both $\bs K$ and $\bs L$ by the superdifferential operators
 $[\rD_{[3}\rD_{4]}-\rD_{[1}\rD_{2]}]$ and $[\rD_{[1}\rD_{4]}-\rD_{[2}\rD_{3]}]$|in addition to the imposition of\eq{e:CLSprimary}.

Indeed, we have the iteration of superdifferential constraints
\begin{equation}
  \IT_x \oplus \IT_y =
  \bigg\{\,  (C^{ab} + (\gamma^5)^{ab}){\rm D}_a {\rm D}_b (\bs K + i \bs L) = 0
  \Big\}
            ~\& \begin{array}{rl}
                  [\rD_{[3}\rD_{4]}-\rD_{[1}\rD_{2]}](\bs K+i\bs L)&=0\\[0mm]
                  [\rD_{[1}\rD_{4]}-\rD_{[2}\rD_{3]}](\bs K + i\bs L)&=0
                \end{array}\,\bigg\}.
 \label{e:TxTy}
\end{equation}
Since $[\rD_{[3}\rD_{4]}-\rD_{[1}\rD_{2]}]$ and $[\rD_{[1}\rD_{4]}-\rD_{[2}\rD_{3]}]$ are real operators, they apply {\em\/separately\/} on $\bs K$ and $\bs L$, as desired, thus setting all four bosons $\eX_{34},\eX_{14},\eY_{34}$ and $\eY_{14}$ to zero. Since the supersymmetry transformation of these component fields consists|after the imposition of \Eq{e:Lambdasubs}|purely of $\Tw{\bs \beta}_I$, then supersymmetry implies that the $2^\text{nd}$ ``tier'' of superdifferential constraints\eq{e:TxTy},
\begin{equation}
 [\rD_{[3}\rD_{4]}-\rD_{[1}\rD_{2]}](\bs K+i\bs L)=0 \qquad\text{and}\qquad
 [\rD_{[1}\rD_{4]}-\rD_{[2}\rD_{3]}](\bs K+i\bs L)=0,
 \label{e:2ndT}
\end{equation}
also annihilates $\Tw{\bs \beta}_I$.

In fact, the superdifferential operators $[\rD_{[3}\rD_{4]}-\rD_{[1}\rD_{2]}]$ and $[\rD_{[1}\rD_{4]}-\rD_{[2}\rD_{3]}]$ appear, at least in the $0$-brane reduction, in the constraints that define the \CLS\ itself. The operator $[\rD_{[3}\rD_{4]}-\rD_{[1}\rD_{2]}]$ appears in the primary superdifferential constraint~\eq{e:CLSprimaryRealImaginary}. The $0$-brane reductions, Eq.~\eq{e:DDMN0},
\begin{subequations}\label{e:DDMN0}
\begin{align}
 H \left[\rD_{[1} \rD_{4]} - \rD_{[2} \rD_{3]} \right] \bs K &= - 2 \left[  H^2 + {\rm D}_1 {\rm D}_2 {\rm D}_3 {\rm D}_4 \right]\bs L \\
2 \left[ H^2 + {\rm D}_1 {\rm D}_2 {\rm D}_3 {\rm D}_4 \right] \bs K &= H \left[\rD_{[1} \rD_{4]} - \rD_{[2} \rD_{3]} \right] \bs L
\end{align}
\end{subequations} of the secondary superdifferential constraints\eq{e:DDMN} contains the superdifferential operator $[\rD_{[1}\rD_{4]}-\rD_{[2}\rD_{3]}]$. However, therein, the $[\rD_{[3}\rD_{4]}-\rD_{[1}\rD_{2]}]$ and $[\rD_{[1}\rD_{4]}-\rD_{[2}\rD_{3]}]$ occur only:
\begin{enumerate}\itemsep=-3pt\vspace{-2mm}
 \item on one side of the original \CLS\ superdifferential constraints\eq{e:CLSprimaryRealImaginary}, and
 \item on one side of the $0$-brane reduced secondary \CLS\ superdifferential constraints \eq{e:DDMN0}.
\end{enumerate}
Therefore, imposing\eq{e:2ndT} in addition to\eq{e:CLSprimaryRealImaginary} sets {\em\/both\/} sides of the equations\eq{e:CLSprimaryRealImaginary} and \eq{e:DDMN0} to zero {\em\/separately\/}. This makes it obvious that $\IT_x\oplus\IT_y$, as defined in\eq{e:TxTy}, is a special case of \CLS, and therefore a sub-supermultiplet,
\begin{equation}
  \IT_x\oplus\IT_y\subset\CLS,
  \qquad\text{and then}\qquad
 \CLS_V=\CLS/(\IT_x\oplus\IT_y).
% \label{e:}
\end{equation}

Thus, we conclude:
\begin{itemize}\itemsep=-3pt\vspace*{-3mm}
 \item 
The \CLS\ is indecomposable as a $(1|4)$-supermultiplet, but reduces to the direct sum of two smaller $(1|4)$-supermultiplets, $\IT_x\oplus\IT_y\subset\CLS$ upon the imposition of the additional superdifferential constraints\eq{e:2ndT}.

 \item
The \CLS\ is not adinkraic and it does have uni-directional supersymmetry action. The latter is arguably correlated with the reducibility discussed above: by annihilating the $\CLS_V$ ``kite,'' both the remainder of \CLS\ decomposes and the uni-directional $\rD$- ($Q$-)action is eliminated.
\end{itemize}

\clearpage
\section{Actions, Variations, and Conclusions}
 \label{s:End}
 \subsection{Real Scalar Supermultiplet}
 The $4D$ transformation laws for the real scalar supermultiplet\eq{e:DRSM} are an invariance of the action
 \begin{equation}\label{e:LRSS}
   {\mathcal L}_{\text{\RSS}} = -\frc{1}2 \bs M^2 - \frc{1}2\bs N^2 + \frc{1}2 \bs U_\mu \bs U^\mu - \frc{1}2 \bs K {\bs{\rm d}} + i \frc{1}2\bs \zeta_a C^{ab} \bs \Lambda_b 
\end{equation}
up to total derivatives. Reduced to the 0-brane, this action is unchanged and we now write it in terms of the $\bs X_{IJ}$ defined in Eq.\eq{e:RSSX}
\begin{align}\label{e:LRSS0}
    {\mathcal L}_{\text{\RSS}} =& \frc{1}4\epsilon^{IJKL}\bs X_{IJ} \bs X_{KL} - \frc{1}2 \bs K {\bs {\rm d}} +  i \frc{1}2\bs \zeta_a C^{ab} \bs \Lambda_b \cr 
    =& \bs X_{12} \bs X_{34} + \bs X_{31} \bs X_{24} + \bs X_{14} \bs X_{23} - \frc{1}2 \bs K \bs {\rm d} +i \frc{1}2\left(-\bs \zeta_1 \bs \Lambda_2 +\bs \zeta_2 \bs \Lambda_1 + \bs \zeta_3 \bs \Lambda_4 - \bs \zeta_4 \bs \Lambda_3 \right)~~~, \\
    \epsilon^{1234} =& 1~~~\mbox{and totally antisymmetric.}
\end{align}
The Lagrangian~\eq{e:LRSS0} is clearly invariant up to total derivatives with respect to the 0-brane transformations in Table~\ref{t:ScX}.

 \subsection{Real Pseudoscalar Supermultiplet}
 Performing the substitutions~\eq{substitution1} on the \RSS\ action~\eq{e:LRSS} we arrive at the Lagrangian for the real pseudoscalar supermultiplet
 \begin{align}
    {\mathcal L}_{\text{\RPS}} = -\frc{1}2 \Tw{\bs M}^2 - \frc{1}2\Tw{\bs N}^2 + \frc{1}2 \bs V_\mu \bs V^\mu - \frc{1}2 \bs L \Tw{\bs{\rm d}} + i \frc{1}2\bs \rho_a C^{ab} \Tw{\bs \Lambda}_b 
 \end{align}
 which is invariant up to total derivatives with respect to the transformation laws~\eq{e:DRPM}. This action is obviously unchanged upon 0-brane reduction and here we write it in terms of the $\bs Y_{IJ}$ defined in Eq.\eq{e:RPSY}
 \begin{align}\label{e:LRPS0}
    {\mathcal L}_{\text{\RPS}} =& \frc{1}4\epsilon^{IJKL}\bs Y_{IJ} \bs Y_{KL} - \frc{1}2 \bs L \Tw{\bs {\rm d}} +  i \frc{1}2\bs \rho_a C^{ab} \Tw{\bs \Lambda}_b \cr 
    =& \bs Y_{12} \bs Y_{34} + \bs Y_{31} \bs Y_{24} + \bs Y_{14} \bs Y_{23} - \frc{1}2 \bs K \bs {\rm d} +i \frc{1}2\left(-\bs \rho_1 \Tw{\bs \Lambda}_2 +\bs \rho_2 \Tw{\bs \Lambda}_1 + \bs \rho_3 \Tw{\bs \Lambda}_4 - \bs \rho_4 \Tw{\bs \Lambda}_3 \right)~~~
\end{align}
which is invariant up to total derivatives with respect to the transformation laws in Table~\ref{t:ScY}.
 \subsection{Complex Linear Supermultiplet}
 The transformation laws~\eq{e:PreB} are an invariant of the \CLS\ Lagrangian~\eq{e:LCLSPB}
 \begin{align}\label{e:LCLSPB}
 \mathcal{L}_{\text{\CLS}} =& -\frc{1}4\partial_\mu \bs K \partial^\mu \bs K - \frc{1}4 \partial_\mu \bs L \partial^\mu \bs L - \frc{1}2 \bs M^2 - \frc{1}2 \bs N^2 + \cr
 &+ \frc{1}4 \bs U_\mu \bs U^\mu  +\frc{1}4\bs V_\mu \bs V^\mu +\frc{1}2 \bs V^{\mu} \partial_\mu \bs K  - \frc{1}2 \bs U^\mu \partial_\mu \bs L\cr
& + \frc{i}4 (\gamma^\mu)^{ab} \bs \zeta_a \partial_\mu \bs \zeta_b + \frc{i}4  (\gamma^\mu)^{ab} \bs \rho_a \partial_\mu \bs \rho_b + \frc{i}4 (\bs \rho_a + \bs \zeta_a) C^{ab} \bs\beta_b 
 \end{align}
 up to total derivatives. We can remove the cross terms 
 \begin{align}
    \bs V^\mu \partial_\mu \bs K ~~~,~~~\bs U^\mu \partial_\mu \bs L \bs ~~~,~~~ i \bs \zeta_a C^{ab} \bs \beta_b
\end{align}
and the extra kinetic term
\begin{equation}
    i (\gamma^\mu)^{ab} \bs \rho_a \partial_\mu  \bs \rho_b
\end{equation}
after switching to the basis
\begin{subequations}\label{e:CLSshift}
  \begin{align}
     \hat{\bs V}_\mu \equiv & \bs V_\mu + \partial_\mu {\bs K} ~~~,~~~\hat{\bs U}_\mu \equiv \bs U_\mu - \partial_\mu {\bs L} \\
     2 \hat{\bs \rho}_a \equiv & \bs \rho_a + \bs \zeta_a ~~~,~~~ 2 \hat{\bs \zeta}_a \equiv  \bs \rho_a - 
     {\bs \zeta}_a \\
     \hat{\bs \beta}_a \equiv & \frc{1}2\bs \beta_a + \frc{1}4 (\gamma^\mu)_{a}^{~b}\partial_\mu (\bs \rho_b \bs + \zeta_b)
  \end{align}
  \end{subequations}
  with which the \CLS\ Lagrangian takes, up to total derivatives, the simpler form in~\eq{e:LagrangianCLM}
  \begin{align}\label{e:LagrangianCLM}
\mathcal{L}_{\text{\CLS}} &= -\frc{1}2\partial_\mu \bs K \partial^\mu \bs K - \frc{1}2 \partial_\mu \bs L \partial^\mu \bs L - \frc{1}2 \bs M^2 - \frc{1}2 \bs N^2 +\frc{1}4 \hat{\bs U}_\mu \hat{\bs U}^\mu  +\frc{1}4\hat{\bs V}_\mu \hat{\bs V}^\mu \cr
& +\frc{1}2i (\gamma^\mu)^{ab} \hat{\bs \zeta}_a \partial_\mu \hat{\bs \zeta}_b  + i\hat{\bs \rho}_a C^{ab} \hat{\bs\beta}_b ~~~.
\end{align}
The $0$-brane reduction of the original \CLS\ Lagrangian~\eq{e:LCLSPB} in the original, un-hatted basis takes the form in Eq.~\eq{e:LCL0}.
\begin{align}\label{e:LCL0}
   {\mathcal L}_\text{\CLS} = &\frc{1}4 \Dt{\bs K}^2 + \frc{1}4 \Dt{\bs L}^2 - \frc{1}4 \eY_{14}\Dt{\bs K} + \frc{1}4 \eX_{14} \Dt{\bs L} + \frc{1}4\left( \bX_{23}^2 + \bX_{12}^2 + \bX_{24}^2 + \bY_{23}^2 + \bY_{12}^2 + \bY_{24}^2 \right) + \cr &+\frc{1}4\left(\eX_{14}\bX_{23} + \eX_{34} \bX_{12} + \eY_{14}\bY_{23} + \eY_{34} \bY_{23} + \eY_{34} \bY_{12} + \eY_{34} \bX_{24} - \eX_{34}\bY_{24}\right) + \cr
    &+ \frc{i}4\delta^{IJ}\left(\bs \zeta_I \Dt{\bs \zeta_J} +  \bs \rho_I \Dt{\bs \rho_J} \right) + \frc{i}4\left( \bs \rho_{[2}\bs \beta_{1]} + \bs \zeta_{[2}\bs \beta_{1]} + \bs \rho_{[3}\bs \beta_{4]} + \bs \zeta_{[3}\bs \beta_{4]} \right)
\end{align}
The field redefinitions~\eq{e:CLSshift} turn the transformation laws~\eq{e:PreB} into the transformation laws~\eq{e:M4CLS} for the hatted basis 
  \begin{subequations}
\label{e:M4CLS}
\begin{align}
 \rD_a\,\bs K &= \hat{\bs\ro}_a - \hat{\bs\z}_a \\[1mm]
 \rD_a\,\bs L &= i(\g^5)_a^{~b} (\hat{\bs\ro}_b + \hat{\bs\z}_b) \\[2mm]
 \rD_a\,\hat{\bs\ro}_b &= i C_{ab}\,\bs M  + (\g^5)_{a b}\,\bs N
                +\inv2( \g^5 \g^\m)_{ab}\,\hat{\bs U}_\m +\frc{i}2(\g^\m)_{ab}\,\hat{\bs V}\!_\m \\[1mm]
 \rD_a\,\hat{\bs\z}_b &= -i(\g^\m)_{ab}\,(\vd_\m\bs K) - (\g^5 \g^\m)_{ab}\,(\vd_\m\bs L)
              -\inv2( \g^5 \g^\mu)_{ab}\,\hat{\bs U}_\mu + \frc{i}2(\g^\mu)_{ab}\,\hat{\bs V}\!_\mu\\[2mm]
 \rD_a\,\bs M &= \hat{\bs\b}_a
                 -\inv2 (\g^\m)_a{}^b\,(\vd_\m \hat{\bs\ro}_b)\\[0mm]
 \rD_a\,\bs N &=  -i (\g^5)_a{}^b\,\hat{\bs\b}_b
                  +\frc{i}2 (\g^5 \g^\m)_a{}^b\,(\vd_\m\hat{\bs\ro}_b)\\[0mm]
 \rD_a\,\hat{\bs U}_\mu &= i(\g^5 \g_\mu)_a{}^b\,\hat{\bs\b}_b
                    - i (\g^5)_a{}^b\big(\vd_\m(\hat{\bs\ro}_b + 2\hat{\bs\z}_b)\big)
                    -\frc{i}2 (\g^5 \g^\n\g_\m)_a{}^b\,\big(\vd_\n(\hat{\bs\ro}_b - 2\hat{\bs\z}_b)\big)
                    \\[1mm]
 \rD_a\,\hat{\bs V}_\mu &= - (\g_\mu)_a{}^b\,\hat{\bs\b}_b
                     +\big(\vd_\m(\hat{\bs\ro}_a - 2\hat{\bs\z}_a)\big)
                     +\inv2(\g^\n\g_\m)_a{}^b\,\big(\vd_\n(\hat{\bs\ro}_b + 2\hat{\bs\z}_b)\big) \\[2mm]
 \rD_a\,\hat{\bs\b}_b &= \frc{i}2 (\g^\m)_{ab}\,(\vd_\m\bs M)
                  + \inv2 (\g^5\g^\m)_{ab}\,(\vd_\m\bs N)\cr
                &\quad
                  + \inv2 (\g^5\g^\m\g^\n)_{ab}\,(\vd_\m\hat{\bs U}_\n)
                   + \inv4 (\g^5\g^\n\g^\m)_{ab}\,(\vd_\m\hat{\bs U}_\n) \cr 
                &\quad
                    + \frc{i}2 (\g^\m\g^\n)_{ab}\,(\vd_\m\hat{\bs V}\!_\n)
                     + \frc{i}{4} (\g^\n\g^\m)_{ab}\,(\vd_\m\hat{\bs V}\!_\n) \cr
                &\quad
                     +\h^{\m\n}\vd_\m\partial_\n(-iC_{ab}\,\bs K +(\g^5)_{ab}\,\bs L),
\end{align}
\end{subequations}
which are an invariant of the $4D$ \CLS\ Lagrangian~\eq{e:LagrangianCLM}, up to total derivatives. In this hatted basis, the $0$-brane reduced Lagrangian takes the most simple form in Eq.~\eq{e:LCLS0diag}
\begin{align}\label{e:LCLS0diag}
   {\mathcal L}_\text{\CLS} =&\frc{1}2 \Dt{\bs K}^2 + \frc{1}2 \Dt{\bs L}^2  - \frc{1}2 \bs M^2 - \frc{1}2 \bs N^2 +\frc{1}4 \hat{\bs U}_\mu \hat{\bs U}^\mu  +\frc{1}4\hat{\bs V}_\mu \hat{\bs V}^\mu + \frc{i}{2}\delta^{IJ}\hat{\bs \zeta}_I \Dt{\hat{\bs \zeta}}_J + i\left( \hat{\bs \rho}_{[2} \hat{\bs \beta}_{1]} + \hat{\bs \rho}_{[3} \hat{\bs \beta}_{4]}\right)\cr
    =&\frc{1}4 \Dt{\bs K}^2 + \frc{1}4 \Dt{\bs L}^2 - \frc{1}4 \eY_{14}\Dt{\bs K} + \frc{1}4 \eX_{14} \Dt{\bs L} + \frc{1}4\left( \bX_{23}^2 + \bX_{12}^2 + \bX_{24}^2 + \bY_{23}^2 + \bY_{12}^2 + \bY_{24}^2 \right) + \cr &+\frc{1}4\left(\eX_{14}\bX_{23} + \eX_{34} \bX_{12} + \eY_{14}\bY_{23} + \eY_{34} \bY_{23} + \eY_{34} \bY_{12} + \eY_{34} \bX_{24} - \eX_{34}\bY_{24}\right) + \cr
    &+ \frc{i}{2}\delta^{IJ}\hat{\bs \zeta}_I \Dt{\hat{\bs \zeta}}_J + i\left( \hat{\bs \rho}_{[2} \hat{\bs \beta}_{1]} + \hat{\bs \rho}_{[3} \hat{\bs \beta}_{4]}\right)
\end{align}

\subsection{Conclusions}
In this work, we have concentrated on the structure of Adinkra graphs that are inherent
in the complex linear superfield  \CLS. Often an overlooked representation in the 
study of $4D$,  ${\mathcal N} =1$ supersymmetry, its existence has mostly been viewed
as a curiosity that warranted little notice.  We have performed a complete analysis of
this multiplet in a self-contained explanation of how it is obtained by imposing appropriate
super-differential equations on a real scalar and real pseudoscalar superfield 
in the $4D$,  ${\mathcal N} =1$ supersymmetry.  We have shown that under a `0-brane
reduction,' this supermultiplet produces a quasi-Adinkra which demonstrates non-Adinkraicity
with respect the
$N$-cubical or projected $N$-cubical
Adinkras studied previously by the
Doran {\em\/et al\/}.\ collaboration of Refs.\cite{r6-1,r6--1,r6-3,r6-3.2,r6-1.2}.

An important implication of this work is that a large class of representations of supersymmetry
is provided by Adinkra-like diagrams that are {\em {not}}
$N$-cubical.  In fact, the construction given
in section two of this work (beginning with two 
$N$-cubical Adinkras for the real scalar and real 
pseudoscalar supermultiplets) which imposes constraints leads to the non-$N$-cubical Adinkra 
of the complex linear supermultiplet.  In fact, rather than being exceptional, it is more likely 
that as more complicated supersymmetric systems are considered, the occurrence of non-$N$-cubical
Adinkras will become generic.  This has implications for the construction of higher dimensional 
off-shell supermultiplets.  Any study based {\em {solely}} on 
$N$-cubical Adinkras is likely to miss 
the important class of non-cubical representations.   At a minimum such a study would be 
incomplete and inconclusive.

%\clearpage

%\section{Conclusions}
% \label{s:End}
% %
%Wherein we summarize what we just wrote.

\vspace{.05in}
\begin{center}
 \parbox{4in}{{\it ``A fear of the unknown keeps a lot of people from leaving bad situations.''}\,\,-\,\, Kathy Lee Gifford}
 \end{center}   
 
%optional other quotes:
% \vspace{.05in}
% \begin{center}\color{Turque}{
% \parbox{4in}{{\it ``All we know is still infinitely less than all that remains unknown.''}\,\,-\,\, William Harvey}
% \end{center}   
%or
% \vspace{.05in}
% \begin{center}\color{Turque}{
% \parbox{4in}{{\it ``Fear of the unknown is a terrible fear.''}\,\,-\,\, Joan D. Vinge}
% \end{center}   
 
\bigskip\bigskip
\paragraph{\bfseries Acknowledgments:}
We thank Sergei M.~Kuzenko, Ulf Lindstr\" om, Martin Ro\v cek, and Warren Siegel for discussions.
 SJG's and KS's research was supported in part by the endowment of the John S.~Toll Professorship, the University of Maryland Center for String \& Particle Theory, National Science Foundation Grant PHY-0354401. SJG's work is also supported by U.S. Department of Energy (D.O.E.) under cooperative agreement DE-FG02-5ER-41360.  SJG and KS offer additional gratitude to the M.~L.~K.\ Visiting Professorship and to the M.~I.~T.\ Center for Theoretical Physics for support and hospitality extended during the undertaking of this work.
 TH is grateful to the Department of Energy for the generous support through the grant DE-FG02-94ER-40854, as well as the Physics Department of the Faculty of Natural Sciences of the University of Novi Sad, Serbia, for recurring hospitality and resources.
% Some Adinkras were drawn with the help of the \textsl{Adinkramat}~\copyright\,2008 by G.~Landweber.
%\vfill

\newpage\raggedright
\bibliographystyle{elsart-numX}
\bibliography{Refs}

\end{document}